\documentstyle[12pt,psfig]{article}
\newcommand{\mysection}{\setcounter{equation}{0}\section}

\def\beq{\begin{equation}}
\def\eeq{\end{equation}}
\def\beqa{\begin{eqnarray}}
\def\eeqa{\end{eqnarray}}
\def\a {{\rm f}}
\def\T{T}
\def\U{U} 
\def\blocA{\Gamma_{3 \times 3}}

\def\bc{\begin{center}}
\def\ec{\end{center}}

\begin{document}
\begin {flushright}
FSU-HEP-990226\\
\end {flushright} 
\vspace{3mm}
\begin{center}
{\Large \bf Resummation for heavy quark \\
and jet cross sections} 
\end{center}
\vspace{2mm}
\begin{center}
Nikolaos Kidonakis\\
\vspace{2mm}
{\it Department of Physics\\
Florida State University\\
Tallahassee, FL 32306-4350, USA} \\
\end{center}

\begin{abstract}
We review the resummation of threshold logarithms for heavy
quark, dijet, direct photon, and $W$ boson production cross sections 
in hadronic collisions. 
Beyond leading logarithms the resummed cross section
is sensitive to the color exchange in the hard scattering.
The resummation is formulated at next-to-leading logarithmic or higher
accuracy in terms of anomalous dimension matrices which describe the
factorization of soft gluons from the hard scattering.
We give results for the soft anomalous dimension matrices
at one loop for the full range  of partonic subprocesses involved in 
heavy quark, dijet, direct photon, and $W$ boson production. 
We discuss the general diagonalization procedure that can be
implemented for the calculation of the resummed cross sections, 
and we give numerical results for top quark production at the
Fermilab Tevatron. We also present analytical results for the 
one- and two-loop expansions of the resummed cross sections.
 
\end{abstract}
\pagebreak

\mysection{Introduction}

The calculation of a large number of hadronic cross sections in
perturbative Quantum Chromodynamics (QCD) 
relies on factorization theorems~\cite{CSSfact,CoSt,BCSS, CSSn} 
which separate the short-distance hard scattering, calculable in 
perturbation theory, from universal parton
distribution functions that are determined from experiment. 
Factorization introduces a scale $\mu$ which separates the short- and
long-distance physics; the factorization scale is often taken to be
the same as the renormalization scale, though in principle these two
quantities are independent. Leading order calculations for the
cross sections exhibit a strong dependence on $\mu$, but next-to-leading
order (NLO) calculations diminish this scale dependence, and sensitivity
to even higher orders is normally exhibited by the variation of the scale.
NLO results exist for a wide range
of processes including Drell-Yan~\cite{dy1loop}, 
direct-photon~\cite{dgamma1loop}, electroweak boson~\cite{wjet}, 
Higgs~\cite{Higgs},
heavy quark~\cite{heavycalcs}, and jet production~\cite{jet1loop}. 
Results exist at two-loops or higher for some processes such as
Drell-Yan production~\cite{DY2l}, 
$e^+ e^-$ annihilation~\cite{ee2l}, and deep inelastic 
scattering~\cite{dis2l}. Calculations to even higher orders
are formidable and in general one cannot make more precise
predictions. Near threshold for the production of the observed final state,
however, there are large logarithmic corrections which arise from
soft gluon emission and which can be resummed to all orders in 
perturbation theory~\cite{DYearly}. At threshold there is no phase 
space left for the emission of gluonic radiation and the cancellation of
infrared divergences between virtual and real diagrams is incomplete. 
The resultant singular distributions were first resummed for the Drell-Yan 
process~\cite{St87,CT1}; this resummation is related to an earlier
study of the transverse momentum distribution of hadron pairs in
$e^+ e^-$ annihilation~\cite{CoSo81}. 
Here we will review the threshold resummation of Sudakov logarithms 
in heavy quark, dijet, direct photon, and $W$ boson production 
in hadronic collisions, 
giving special attention to the resummation of next-to-leading logarithms. 

The resummation of the leading theshold logarithms in heavy quark
production cross sections~\cite{LSN, pp, HERAB, Thesis, BC, CMNT, SIO} 
and inclusive differential distributions~\cite{pty,pp,Thesis} is 
based on the analogy with the Drell-Yan process 
since the leading logarithms (LL) are universal. 
The top quark has finally been found at the Fermilab Tevatron \cite{CDF,D0}
(for a review see~\cite{toprev1,toprev2}). The top's 
mass and production cross section are
of great interest, and the precision of their 
measurements will increase in future runs, 
so it is important to have good theoretical predictions
for the cross section in terms of the top mass; threshold effects 
are expected to play a significant role in these predictions. 
Threshold resummation will also be of relevance to bottom quark production
at the HERA-B experiment at DESY.

A study of subleading logarithms in heavy quark production near threshold
and an attempt to resum them was first presented in Ref.~\cite{pp}.
The resummation of next-to-leading logarithms (NLL) 
for QCD hard scattering and heavy quark production,
in particular, was first put on a solid theoretical framework
in~\cite{Thesis,NKGS,KS}. It was found that at NLL one has to take into
account the color exchange in the hard scattering.
The resummation of the cross section is achieved by refactorizing
the cross section into hard components associated with the hard scattering,
center-of-mass distributions for the colliding partons, and a soft function
that describes coherent~\cite{Coherence} non-collinear soft 
gluon emission. The soft function is a matrix in
color space and it satisfies a renormalization group equation in terms of
soft anomalous dimensions. The resummation is then given in terms
of the angular-dependent eigenvectors and eigenvalues of the anomalous 
dimension matrix.    
Applications to top and bottom quark production 
at a fixed center-of-mass scattering angle,
$\theta=90^{\circ}$ (where the anomalous dimension matrix is diagonal), 
were discussed in Ref. \cite{NKJSRV}. More recently the 
full angle-integrated cross section was presented and numerical results
given for top quark production at the Tevatron~\cite {NKRV}.
A different formalism~\cite{BCMN} which avoids the angular dependence
in the resummed corrections has appeared recently for the 
heavy quark total cross section only (and also for direct photon
production~\cite{CMN}).
The formalism of Refs.~\cite{Thesis,NKGS,KS} and its extensions
that we review in this paper is 
more general since it keeps an explicit angular dependence; it can thus also 
be applied easily to differential distributions~\cite{KLMV}.

Natural extensions of the NLL resummation formalism to the
hadroproduction of dijets have been presented in Refs.~\cite{KOS1,KOS2}.
In addition, rapidity gaps in high-$p_T$ dijet events have been
studied by resumming logarithms of the energy flow into the central
rapidity interval~\cite{OS}.
The resummation formalism has also been applied to single-particle inclusive
cross sections~\cite{LOS} (including direct photon production)
and the electroproduction of heavy quarks~\cite{LM}.
Extensions to $W$ + jet production are also straightforward~\cite{NKVD}.

The NLL resummation formalism that we review here concerns 
hadronic processes of the form
\beq
h_a(p_a)+h_b(p_b) \rightarrow T(Q^2,y,\chi)+X \, ,
\eeq
where $h_a$, $h_b$, are incoming hadrons and $T(Q^2,y,\chi)$ denotes  
the hadronic final state, such as a pair of heavy quarks
or high-$p_T$ jets, produced with large invariant mass $Q$ and rapidity $y$. 
The variable $\chi$
represents the internal structure of the final state, for example
the angle $\theta$ between the direction of the produced 
heavy quark, or jet, and the beam axis, or the rapidity interval
between the two final-state jets.
The cross section for such processes 
may be written in factorized form as a convolution
of the perturbatively calculable 
hard-scattering partonic cross section, $H_{f_a f_b}$,
with parton distributions $\phi_{f_i/h}$, at factorization scale $\mu$,
for parton $f_i$ carrying a momentum fraction $x_i$ of hadron $h$.
Thus, we have: 
\beqa
\frac{d\sigma_{h_a h_b\rightarrow T}(S,Q^2,y,\chi)}{dQ^2 \; dy \; d\chi}&=&
\sum_{f_a,f_b} \; 
\int \frac{dx_a}{x_a} \, \frac{dx_b}{x_b} \,  \phi_{f_a/h_a}(x_a,\mu^2) \, 
\phi_{f_b/h_b}(x_b,\mu^2)
\nonumber \\ && \quad
\times \, H_{f_af_b}\left(\frac{Q^2}{x_a x_b S},y,\chi,{Q\over \mu},
\alpha_s(\mu^2)\right) \, ,
\label{convolution}
\eeqa
where $S$ is the center-of-mass (c.m.) energy squared of the incoming 
hadrons and we sum over all relevant partons $f_a,f_b$.
Although the physical cross section cannot depend on the choice of
factorization scheme (usually DIS or $\overline{\rm MS}$)
and factorization scale $\mu$, at any fixed order there is such
a dependence.  
$H_{f_a f_b}$ includes singular distributions which arise,
as we mentioned before, from incomplete  
cancellations between cross sections with gluon emission and with
virtual gluon corrections.  
These are ``plus'' distributions, of the form 
$[\ln^m(1-z)/(1-z)]_+$, $m \le 2n-1$ at $n$th order in the
perturbative expansion, which are singular for $z=1$, where
\begin{equation}
z=\frac{Q^2}{x_a x_b S}=\frac{Q^2}{s} \, ,
\end{equation}
with $s=x_ax_bS$ the invariant mass squared
of the partons that initiate the hard scattering.  
We shall refer to $z=1$ as  ``partonic threshold'' 
or the ``elastic limit''.
We note that by partonic threshold, we
mean that the c.m. total energy of the incoming partons
is just enough to produce a fixed final state,
such as a pair of heavy quarks or jets; thus, heavy
quarks, for example, are not necessarily produced at rest.

The exponentiation of singular distributions 
is derived in terms of moments of the cross section
with respect to 
\beq
\tau=\frac{Q^2}{S} \, .
\eeq  
Under moments, the convolution in Eq.~(\ref{convolution}) becomes a product
of moments of the parton distributions and the
partonic cross section.
The exponentiated cross section in moment space must
be inverted back to momentum space to derive the physical cross section.
There are a number of different prescriptions (and an ensuing debate in 
the literature) for implementing this inversion and avoiding the Landau pole
that have been proposed for the Drell-Yan process \cite{AMS,CSt,AC} 
and heavy quark production \cite{LSN,BC,CMNT}. These prescriptions do not 
of course exhaust all possibilities. 
Here, we will mostly be concerned with the resummed cross section
in moment space.  

In Section 2 we present the resummation formalism for heavy quark
production. 
In Sections 3 and 4 we give explicit results for the 
soft anomalous dimension matrices in the channels 
$q{\bar q} \rightarrow Q {\bar Q}$ and 
$gg \rightarrow Q {\bar Q}$, relevant to heavy quark production.
Some technical details of the calculation are given in the Appendix.
We also give one- and two-loop expansions of the resummed cross section.
The anomalous dimension matrices are in general not diagonal.
In Section 5 we discuss the diagonalization procedure and present
some numerical results for top quark production at the Fermilab Tevatron.
In Section~6 we discuss threshold resummation for dijet production
and consider the additional complications due to the final state jets.
In Sections 7, 8, and 9, we give results for the anomalous dimension matrices 
for the many subprocesses relevant to dijet production. 
In Section 10 we briefly discuss
resummation for direct photon and $W$ boson production in the context of
single-particle inclusive kinematics.
We conclude with a summary.  

\mysection{Threshold resummation for heavy quark 
\protect\newline production}

In this section we review the general formalism for the 
resummation of threshold singularities for
heavy quark production. We first write the 
factorized form of the cross section
and identify singular distributions in it near threshold. Then we
refactorize the cross section into 
functions associated with gluons collinear to the incoming quarks, 
non-collinear soft gluons,
and the hard scattering. Resummation follows from the renormalization
properties of these functions and is given in terms of
anomalous dimension matrices for each partonic subprocess involved. 

\subsection{Factorized cross section}

We consider the production of a pair of heavy quarks of momenta
$p_1$, $p_2$,
in collisions of incoming hadrons $h_a$ and $h_b$ with momenta
$p_a$ and $p_b$, 
\beq
h_a(p_a)+h_b(p_b) \rightarrow {\bar Q}(p_1) +Q(p_2) + X \, ,
\eeq
with total rapidity $y$
and scattering angle $\theta$ in the pair center-of-mass frame.
The production cross section can be written in a factorized form as 
\beqa
\frac{d\sigma_{h_a h_b\rightarrow Q{\bar Q}}(S,Q^2,y,\theta)}
{dQ^2 \; dy \; d\cos\theta}&=&\sum_{f{\bar f}=q{\bar q},gg} \; 
\int \frac{dx_a}{x_a} \, \frac{dx_b}{x_b} \,  \phi_{f/h_a}(x_a,\mu^2) \,
\phi_{{\bar f}/h_b}(x_b,\mu^2)
\nonumber \\ && \quad
\times \, H_{f{\bar f}}\left(\frac{Q^2}{x_a x_b S},y,\theta,{Q\over \mu},
\alpha_s(\mu^2)\right) \, ,
\label{convolutionHQ}
\eeqa
where $S=(p_a+p_b)^2$, $Q^2=(p_1+p_2)^2$, and we sum over the two
main production partonic subprocesses, $q{\bar q} \rightarrow Q {\bar Q}$ 
and $gg \rightarrow Q {\bar Q}$.
The short-distance hard scattering $H_{f {\bar f}}$ is a smooth function
only away from the edges of partonic phase space. The parton
distribution functions are given in terms of the partonic momentum fractions
and the factorization scale $\mu$, which separates long-distance
from short-distance physics. 

By using the observation of \cite{LaSt} that we can treat the total rapidity
of the heavy quark pair as a constant, equal to its value at threshold,
we can rewrite the above cross section as  
\beqa
\frac{d\sigma_{h_a h_b\rightarrow Q{\bar Q}}(S,Q^2,y,\theta)}
{dQ^2 \; dy \; d\cos\theta}
&=& \sum_{f{\bar f}=q{\bar q},gg}\, \int_\tau^1 dz\, 
\int \frac{dx_a}{x_a} \, \frac{dx_b}{x_b} \, 
\phi_{f/h_a}(x_a,\mu^2) 
\nonumber \\ && \times \, 
\phi_{{\bar f}/h_b}(x_b,\mu^2) \, 
\delta\left (z-{Q^2\over x_ax_bS}\right)\,
\delta\left( y-{1\over2}\ln{x_a\over x_b} \right)
\nonumber \\ && \quad \quad \times \,  
{\hat \sigma}_{f{\bar f}\rightarrow Q{\bar Q}}\left(1-z, 
\frac{Q}{\mu},\theta,\alpha_s(\mu^2)\right) \, ,
\eeqa
where in addition we have introduced an explicit integration over $z=Q^2/s$,
and a simplified hard scattering function ${\hat \sigma}$.

In general, ${\hat \sigma}$ includes ``plus''
distributions with respect to $1-z$, 
with singularities at $n$th order in $\alpha_s$ of the type
\beq
-\frac{\alpha_s^n}{n!}\, \left[\frac{\ln^{m}(1-z)}{1-z} \right]_{+}, 
\hspace{10mm} m\le 2n-1\, .
\eeq
These ``plus'' distributions are defined by their integrals with 
any smooth functions ${\cal F}(z)$
(such as parton distribution functions) by
\beqa
\int_y^1 dz \left[ {\ln^{m}(1-z) \over 1-z} \right ]_+\; {\cal F}(z)
&=&
\int_y^1 dz \left[ {\ln^{m}(1-z) \over 1-z} \right ] \left[{\cal F}(z)
{}-{\cal F}(1)\right]\nonumber\\
&\ & \quad - {\cal F}(1)\; \int_0^y dz \left[ {\ln^{m}(1-z) \over 1-z} 
\right ]\, .
\eeqa
All distributions of this kind have been resummed for the Drell-Yan 
production cross section at leading and nonleading
logarithms~\cite{St87,CT1}. Here we will review the more recent work
on resummation for heavy quark production including next-to leading 
logarithms~\cite{Thesis,NKGS,KS}. 
 
In order to calculate the hard-scattering function 
${\hat \sigma}_{f{\bar f}\rightarrow Q{\bar Q}}$,
we consider the infrared regularized parton-parton scattering 
cross section, which factorizes as the hadronic
cross section.
After we integrate over rapidity, we have 
\beqa
\frac{d\sigma_{f{\bar f}\rightarrow Q{\bar Q}}(S,Q^2,\theta,\epsilon)}
{dQ^2 \; d\cos \theta}
&=&  \int_\tau^1 dz \, 
\int \frac {dx_a}{x_a} \, \frac{dx_b}{x_b} \, 
\phi_{f/f}(x_a,\mu^2,\epsilon) \, 
\phi_{{\bar f}/{\bar f}}(x_b,\mu^2,\epsilon)
\nonumber \\ && \hspace{-13mm} \times \, 
\delta\left (z-{Q^2\over x_ax_bS}\right) \,  
{\hat \sigma}_{f{\bar f}\rightarrow Q{\bar Q}}\left(1-z, 
\frac{Q}{\mu},\theta,\alpha_s(\mu^2)\right) \, .
\eeqa
The argument $\epsilon$ represents the universal collinear
singularities.
We note that the leading power as $z \rightarrow 1$ comes
entirely from the flavor diagonal parton distributions
$\phi_{f/f}(x_a,\mu^2,\epsilon)$ and  
$\phi_{{\bar f}/{\bar f}}(x_b,\mu^2,\epsilon)$.

If we take Mellin transforms of the above equation, the convolution
becomes a simple product of the moments of the parton distributions
and the hard scattering function ${\hat \sigma}$:
\beqa
&& \int_0^1 d\tau \tau^{N-1}
\frac{d\sigma_{f{\bar f}\rightarrow Q{\bar Q}}(S, Q^2, \theta, \epsilon)}
{dQ^2 \; d\cos \theta}
\nonumber \\ && \quad  
={\tilde \phi}_{f/f}(N,\mu^2,\epsilon)\; 
{\tilde \phi}_{{\bar f}/{\bar f}}(N,\mu^2,\epsilon) \;
{\hat \sigma}_{f{\bar f}\rightarrow Q{\bar Q}}(N,
Q/\mu,\theta,\alpha_s(\mu^2)) \, ,
\label{sigmom}
\eeqa
with the moments given by 
$\tilde{\sigma}(N)=\int_0^1dz\; z^{N-1}\hat{\sigma}(z)$, 
$\tilde{\phi}(N)=\int_0^1dx\; x^{N-1}\phi(x)$.  
We then factorize the initial-state collinear divergences
into the parton distribution functions, expanded to the same
order in $\alpha_s$ as the partonic cross section, 
and thus obtain the perturbative expansion for the
infrared-safe hard-scattering function, ${\hat \sigma}$.

We note that under moments divergent distributions in $1-z$ produce 
powers of $\ln N$:
\beq
\int_0^1 dz\; z^{N-1}\left[{\ln^m(1-z)\over 1-z}\right]_+
={(-1)^{m+1}\over m+1}\ln^{m+1}N +{\cal O}\left(\ln^{m-1}N\right)\, .
\eeq

The hard scattering function ${\hat \sigma}$ still has sensitivity
to soft-gluon dynamics through its $1-z$ dependence (or the $N$ dependence
of its moments).   
We may now refactorize (moments of) the cross section 
into $N$-independent hard components $H_{IL}$, 
which describe the truly short-distance hard-scattering,
center-of-mass distributions $\psi$, associated with gluons
collinear to the incoming partons, and a soft gluon function
$S_{LI}$ associated with non-collinear soft gluons. $I$ and $L$
are color indices that describe the color structure
of the hard scattering.
Then we write the refactorized cross section as~\cite{KS}
\beqa
&& \hspace{-10mm} \int_0^1 d\tau \tau^{N-1}
\frac{d\sigma_{f{\bar f}\rightarrow Q{\bar Q}}(\tau,Q^2,\theta,\epsilon)}
{dQ^2 \; d\cos \theta}
=\sum_{IL} H_{IL}\left({Q\over\mu},\theta, \alpha_s(\mu^2)\right)\; 
\nonumber \\ && \hspace{-10mm} \times \; 
{\tilde S}_{LI} \left({Q\over N \mu },\theta, \alpha_s(\mu^2) \right)\;
{\tilde\psi}_{f/f}\left ( N,{Q\over \mu },\epsilon \right) \;
{\tilde\psi}_{{\bar f}/{\bar f}}\left (N,{Q\over \mu },\epsilon \right )  
+{\cal O}(1/N) \, .
\label{sigref}
\eeqa
This factorization is illustrated in Fig. 1, for
the process
\beq
f(p_a)+{\bar f}(p_b)\rightarrow {\bar Q}(p_1) + Q(p_2)\, ,
\eeq
in which $f{\bar f}$ represents a pair of light quarks
or gluons that annihilate or fuse, respectively, into a heavy quark pair.

\begin{figure}
\centerline{
\psfig{file=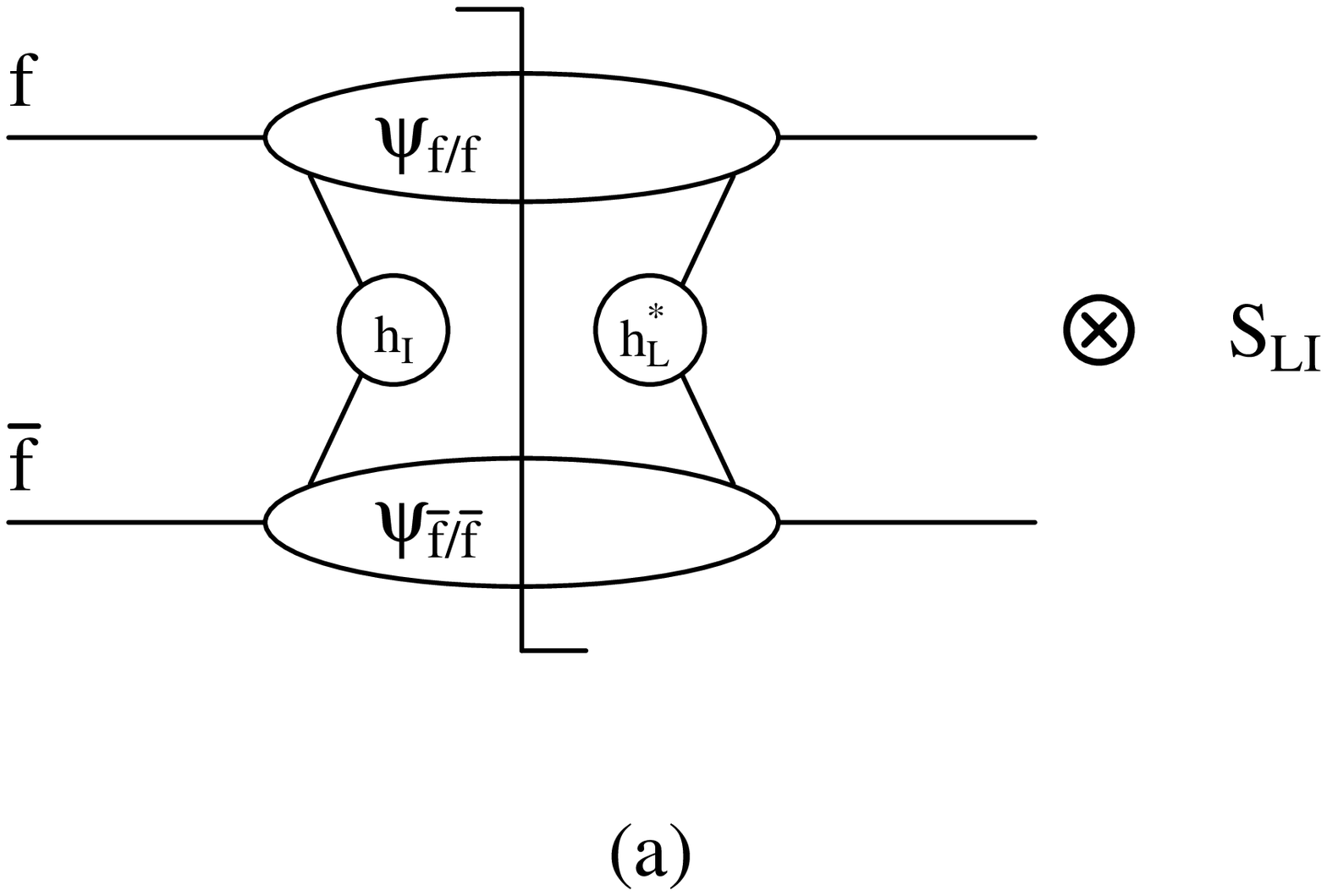,height=2.5in,width=4.05in,clip=}}
\vspace{5mm}
\centerline{
\psfig{file=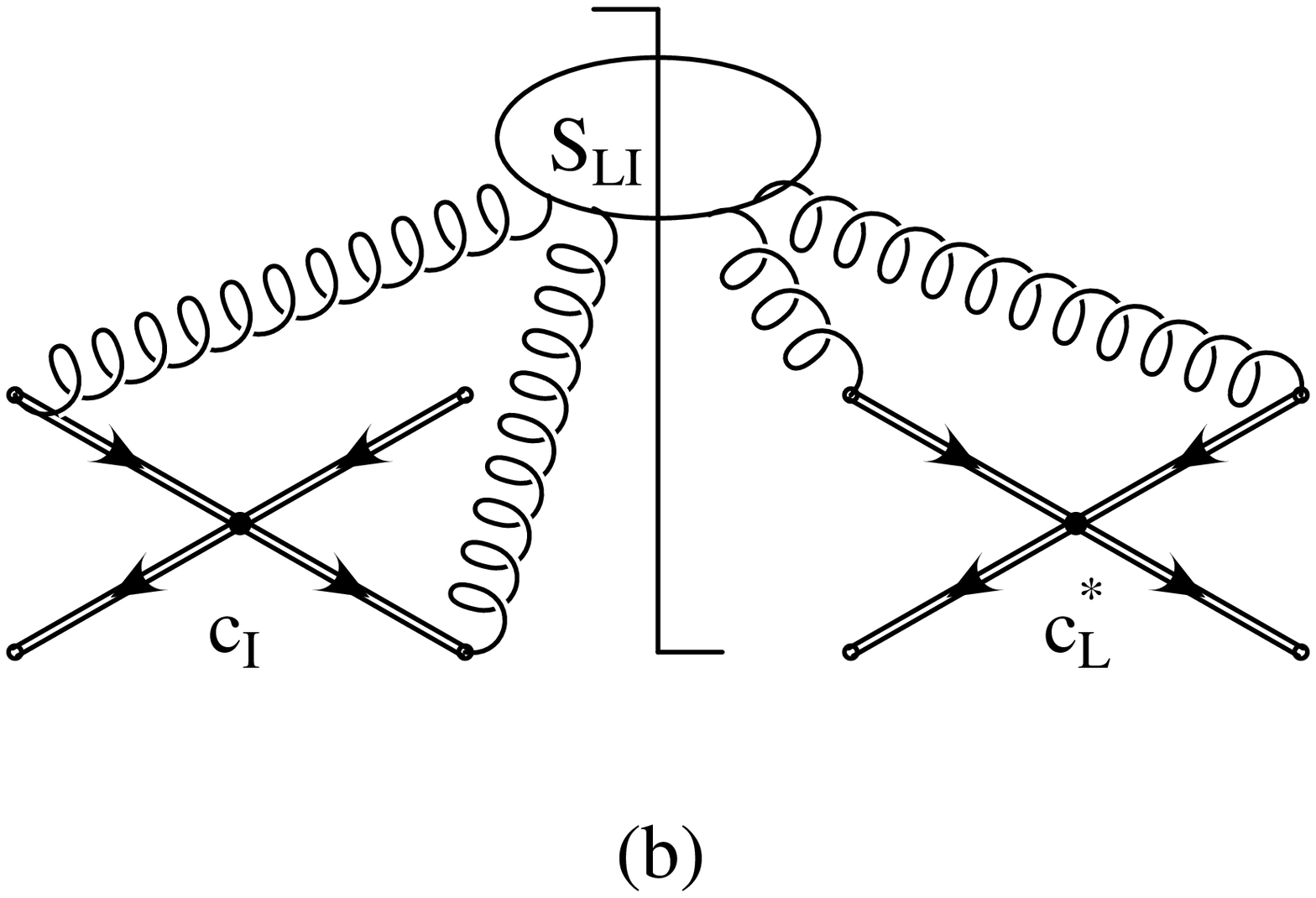,height=2.5in,width=4.05in,clip=}}
{Fig. 1. (a) Factorization for heavy quark production near partonic threshold.
(b) The soft-gluon function $S_{LI}$, in which
the vertices $c_I,c_L^*$ link ordered exponentials.}
\label{fig1}
\end{figure}

The hard-scattering function takes contributions from the 
amplitude and its complex conjugate,
\beq
H_{IL}\left({Q\over\mu},\theta,\alpha_s(\mu^2)\right) 
=
{h^*}_{L}\left({Q\over\mu},\theta,\alpha_s(\mu^2)\right)\;
h_{I}\left({Q\over\mu},\theta,\alpha_s(\mu^2)\right).
\nonumber\\ 
\eeq

The center-of-mass distribution functions $\psi$ absorb 
the universal collinear singularities
associated with the initial-state partons in the refactorized cross section. 
They differ from standard light-cone parton 
distributions by being defined at fixed energy, rather than
light-like momentum fraction.  
We define them by analogy to the light-cone parton distributions $\phi$
via the matrix elements, evaluated in $n\cdot A=0$ axial gauge
in the partonic c.m. frame, 
\beqa
\psi_{q/q}(x,2p_0/\mu,\epsilon)
&=&\frac{1}{2\pi 2^{3/2}}
\int_{-\infty}^\infty dy_0\ {\rm e}^{-ixp_0y_0}
\langle q(p)|\bar{q}(y_0,\vec{0})\frac{1}{2}v\cdot \gamma 
q(0)|q(p)\rangle \, ,
\nonumber \\
\psi_{\bar{q}/\bar{q}}(x,2p_0/\mu,\epsilon)
&=&\frac{1}{2\pi 2^{3/2}}
\nonumber \\ && \times \;
\int_{-\infty}^\infty dy_0\ {\rm e}^{-ixp_0y_0}
\langle \bar{q}(p)|
{\rm Tr}\left[\frac{1}{2}v\cdot \gamma q(y_0,\vec{0})
\bar{q}(0)\right]\ |\bar{q}(p)\rangle \, ,
\nonumber \\
\psi_{g/g}(x,2p_0/\mu,\epsilon)
&=&\frac{1}{2\pi 2^{3/2}p^+}
\nonumber \\ && \times \;
\int_{-\infty}^\infty dy_0\ {\rm e}^{-ixp_0y_0}
\langle g(p)|v_\mu F^{\mu\perp}(y_0,\vec{0}) 
v_\nu {F^\nu}_\perp(0)|g(p)\rangle\, ,
\nonumber\\
\label{psidef}
\eeqa
where the vector $v$ is light-like in the opposite direction from $p^\mu$,
and the argument $\epsilon$ denotes the collinear singularities
that $\psi$ absorbs.
The moments of $\psi$ can then be written as products of moments
of the light-cone parton distributions $\phi$ and an infrared safe function.

\subsection{Eikonal cross section}

The soft function $S_{LI}$ represents the coupling of soft gluons to the
partons in the hard scattering process. This coupling may be
described by ordered exponentials \cite{PolAr}, or eikonal or Wilson lines,
written as \cite{KOS1,KOS2}
\beq                                                   
\Phi_{\beta}^{(f)}({\lambda}_2,{\lambda}_1;x)=
P\exp\left(-ig\int_{{\lambda}_1}^{{\lambda}_2}d{\lambda}\; 
{\beta}{\cdot} A^{(f)} ({\lambda}{\beta}+x)\right)\, ,
\label{ordexp}
\eeq
where  the gauge field $A^{(f)}$ is a matrix
in the representation of the parton flavor $f$ of the gauge group 
$SU(3)$, $\beta$ is the four-velocity of the
corresponding parton, and $P$ is an operator that orders group products
in the same sense as ordering in the variable $\lambda$.

The color tensors of the hard scattering connect
together the eikonal lines to which soft gluons couple.
We can construct an eikonal operator describing 
soft-gluon emission as
\beqa
&& w_I(x)_{\{c_i\}}=
\Phi_{\beta_2}^{(f_2)}(\infty,0;x)_{c_2,d_2}\;
\Phi_{\beta_1}^{(f_1)}(\infty,0;x)_{c_1,d_1}
\nonumber \\ &&  \times
\left(c_I\right)_{d_2d_1,d_bd_a}\;
\Phi_{\beta_a}^{(f_a)}(0,-\infty;x)_{d_a,c_a}
\Phi_{\beta_b}^{(f_b)}(0,-\infty;x)_{d_b,c_b}\, ,
\label{wvert}
\eeqa
with $c_I$ a color tensor.
Then, we may write a dimensionless eikonal cross section,
which describes the emission of soft gluons by the eikonal lines as
\beqa
\sigma^{({\rm eik})}_{LI}
\left(\frac{wQ}{\mu},\theta,\alpha_s(\mu^2),\epsilon\right)
&=&\sum_{\xi} \delta \left(w-w(\xi)\right)
\nonumber \\ && \hspace{-20mm} \times \,
\langle 0|{\bar T}\left[\left(w_L(0)\right)^{\dagger}_{\{c_i\}}\right]|
\xi \rangle \, \langle \xi|
T\left[w_I(0)_{\{c_i\}}\right]|0\rangle\, ,
\label{sigeik}
\eeqa
where $\xi$ is a set of intermediate states which contribute
to the weight, $w$.
The moments of $\sigma^{\rm (eik)}$
can be factorized, like the moments of the full cross section,
into moments of the soft gluon function, $S$, times moments of jet functions,
$j_{\rm IN}$, analogous to the $\psi$'s, which absorb the collinear
divergences of the incoming eikonal lines.
Hence $S$ is free of these collinear divergences. Thus, we may write
\beqa
{\tilde \sigma}^{({\rm eik})}_{LI}
\left(\frac{Q}{N\mu},\theta,\alpha_s(\mu^2),\epsilon\right)
&=&{\tilde S}_{LI}\left({Q\over N\mu},\theta,\alpha_s(\mu^2)\right)
\nonumber\\ && \hspace{-25mm} \times \,
{\tilde j}_{\rm IN}^{(f_a)}
\left({Q\over N\mu},\alpha_s(\mu^2),\epsilon\right)\; 
{\tilde j}_{\rm IN}^{(f_b)}
\left({Q\over N\mu},\alpha_s(\mu^2),\epsilon\right)\, ,
\eeqa
where the incoming eikonal jet functions are given by 
products of eikonal lines as \cite{KS,KOS1}
\beqa
&& \hspace{-20mm} 
j_{\rm IN}^{(f_i)}\left({w_iQ\over \mu},\alpha_s(\mu^2),\epsilon\right) 
={Q\over 2\pi} \int_{-\infty}^\infty dy_0\ {\rm e}^{-iw_iQy_0}
\nonumber \\ && \times \, 
\langle 0|\; {\rm Tr}\bigg\{\; {\bar T}[\Phi_{\beta_i}^{(f_i) \dagger}
(0,-\infty;y)]
\; T[\Phi_{\beta_i}^{(f_i)}(0,-\infty;0)]\; \bigg\}\; |0\rangle\, ,
\label{eikinjet}
\eeqa
with $y^\nu=(y_0,\vec{0})$ a vector at the spatial origin.

\subsection{Resummation}

Now, comparing Eqs.\ (\ref{sigmom}) and (\ref{sigref}), we
see that the moments of the
heavy-quark production cross section are given by 
\beqa
{\hat \sigma}_{f{\bar f}\rightarrow Q{\bar Q}}(N)&=&
\left[{{\tilde\psi}_{f/f}(N,Q/\mu,\epsilon)
\over{\tilde \phi}_{f/f}(N,\mu^2,\epsilon)}\right]^2\,
\sum_{IL} H_{IL}\left(\frac{Q}{\mu},\theta,\alpha_s(\mu^2)\right)
\nonumber \\ && \quad \quad \times \;
{\tilde S}_{LI}\left(\frac{Q}{N\mu},\theta,\alpha_s(\mu^2)\right) \, ,
\label{psiphiHS}
\eeqa
where $f{\bar f}$ denotes $q{\bar q}$ or $gg$,
the sum over $I$ and $L$ is over color tensors,
and we have used the fact that
${\psi}_{q/q}={\psi}_{{\bar q}/{\bar q}}$
and ${\phi}_{q/q}={\phi}_{{\bar q}/{\bar q}}$.
Each of the factors in Eq.~(\ref{psiphiHS}) is gauge and factorization
scale dependent. The constraint that the product of these factors
is independent of the choice of gauge and factorization scale results
in the exponentiation of logarithms of $N$ in $\psi/\phi$ and $S_{LI}$ 
\cite{Gath,CLS}. We now proceed to discuss exponentiation for each factor.

\subsubsection{Parton distributions}

The first factor, $(\psi_{f/f}/\phi_{f/f})^2$, in Eq.~(\ref{psiphiHS})
is ``universal" between electroweak and
QCD-induced hard processes, and was computed first with $f=q$ for
the Drell-Yan cross section \cite{St87}.  
First, let's present the resummed expression for
the ratio $\psi/\phi$, with $\mu=Q$. We have
\beq
\frac{{\tilde{\psi}}_{f/f}(N,1,\epsilon)}
{{\tilde{\phi}}_{f/f}(N,Q^2,\epsilon)}
=R_{(f)}\left(\alpha_s(Q^2)\right)\; \exp \left[E^{(f)}(N,Q^2)
\right]\, ,
\label{psiphi}
\eeq
where
\beqa
E^{(f)}\left(N,Q^2\right)
&=&
-\int^1_0 dz \frac{z^{N-1}-1}{1-z}\; 
\left \{\int^{(1-z)^{m_S}}_{(1-z)^2} \frac{d\lambda}{\lambda} 
A^{(f)}\left[\alpha_s(\lambda Q^2)\right] \right.
\nonumber\\ &&  \hspace{-15mm} \left.
{}+B^{(f)}\left[\alpha_s((1-z)^{m_s} Q^2)\right]
+\frac{1}{2}\nu^{(f)}\left[\alpha_s((1-z)^2 Q^2)\right] \right\}
\label{Eexp}
\eeqa
and
$R_{(f)}(\alpha_s)$ is an $N$-independent function of the coupling, 
which can be  normalized to unity at zeroth order.
The parameter $m_S$ and  the resummed coefficients $B^{(f)}$ depend on the 
factorization scheme. This scheme-dependence
must be compensated for by differences in the 
parton distributions themselves.
In the  DIS and $\overline{\rm MS}$ factorization schemes
we have $m_S=1$ and $m_S=0$, respectively.

$A^{(f)},B^{(f)}$ and $\nu^{(f)}$ are finite functions 
of their arguments and below we give expressions for them needed 
at next-to-leading order accuracy in $\ln N$.
We have \cite{St87,CT1}
\beq
A^{(f)}(\alpha_s) = C_f\left ( {\alpha_s\over \pi} 
+\frac{1}{2} K \left({\alpha_s\over \pi}\right)^2\right )\, ,
\label{Aexp}
\eeq
with $C_f=C_F=(N_c^2-1)/(2N_c)$ for an incoming quark, 
and $C_f=C_A=N_c$ for an incoming gluon, with $N_c$ the number of colors.
Also
\beq
K= C_A\; \left ( {67\over 18}-{\pi^2\over 6 }\right ) - {5\over 9}n_f\, ,
\eeq
where $n_f$ is the number of quark flavors \cite{KoTr}.  
$B^{(f)}$ is given for quarks in the DIS scheme by
\beq
B^{(q)}(\alpha_s)=-{3\over 4}C_F\; {\alpha_s\over\pi}\, , 
\label{Bexp}
\eeq
while it vanishes in the $\overline {\rm MS}$ scheme for quarks and gluons.
Note that the DIS scheme is normally applied to quarks only,
but extended definitions for gluons are possible~\cite{OwensTung}.  
Finally, the lowest-order approximation to  
the scheme-independent $\nu^{(f)}$ is~\cite{KS}
\beq
\nu^{(f)}=2C_f\; {\alpha_s\over\pi}\, .
\label{nuf}
\eeq

The above results were for $\mu=Q$. To change the scale $\mu$, we need the
renormalization group behavior of the parton distributions
$\psi$ and $\phi$.

The distribution $\psi$, and each of its moments, 
renormalizes multiplicatively,
because it is the matrix element of a product of renormalized
operators.  Thus, it obeys the renormalization group equation
\beq
\mu{d\tilde{\psi}_{f/f}(N,Q/\mu,\epsilon) \over d \mu}
=2\gamma_f(\alpha_s(\mu^2))\; \tilde{\psi}_{f/f}(N,Q/\mu,\epsilon)\, ,
\eeq
where $\gamma_f$ is the anomalous dimension of the field of flavor $f$,
which is independent of $N$.  

The evolution of the light-cone distribution $\tilde{\phi}_{f/f}$ with 
the factorization scale $\mu$ depends on the factorization scheme 
that we choose, such as $\overline{\rm MS}$ or DIS. 
The simplest case is the $\overline{\rm MS}$ factorization scheme.  
In this scheme, the moments of $\phi$ obey the renormalization
group equation 
\beq
\mu{d\tilde{\phi}_{f/f}(N,\mu^2,\epsilon) \over d \mu}
=2\gamma_{ff}(N,\alpha_s(\mu^2))\; \tilde{\phi}_{f/f}(N,\mu^2,\epsilon)\, ,
\eeq
where $\gamma_{ff}$ is the anomalous dimension of the color-diagonal
splitting function for flavor $f$~\cite{CoSoPDF}.
Since only color-diagonal splitting functions are singular as 
$x\rightarrow 1$, only the flavor-diagonal evolution survives in the
large $N$ limit.

Then the generalization of  Eq.\ (\ref{psiphi}) is
\beqa
\frac{{\tilde{\psi}}_{f/f}(N,Q/\mu,\epsilon)}
{{\tilde{\phi}}_{f/f}(N,\mu^2,\epsilon)}
&=& R_{(f)}\left(\alpha_s(\mu^2)\right)\; 
\exp \left[E^{(f)}(N,Q^2)\right]
\nonumber\\ && \hspace{-40mm} \times
\exp \left\{-2\int_{\mu}^Q \frac{d\mu'}{\mu'}\; 
\left [\gamma_f(\alpha_s(\mu'{}^2))
-\gamma_{ff}(N,\alpha_s(\mu'{}^2)) \right]\right\}\, .
\label{psiphimu}
\eeqa

\subsubsection{Renormalization of the hard and soft functions}

Next, we discuss resummation for the soft function.
The soft matrix $S_{LI}$ depends on $N$ through the ratio $Q/(N\mu)$,
and it requires renormalization as a composite operator.
Its $N$-dependence, then, can be resummed by renormalization group
analysis~\cite{KoRa,BottsSt,GK,KK}.
However, the product $H_{IL}S_{LI}$
of the soft function and the hard factors needs
no overall renormalization, because the UV divergences of $S_{LI}$
are balanced by construction by those of $H_{IL}$.
Thus, we have~\cite{KS}
\beqa
H^{(0)}_{IL}&=& \prod_{i=a,b} Z_i^{-1}\; \left(Z_S^{-1}\right)_{IC}
H_{CD} \left[\left(Z_S^\dagger \right)^{-1}\right]_{DL} \, ,
\nonumber \\ 
S^{(0)}_{LI}&=&(Z_S^\dagger)_{LB}S_{BA}Z_{S,AI},
\label{HSren}
\eeqa
where $H^{(0)}$ and $S^{(0)}$ denote the unrenormalized quantities,
$Z_i$ is the renormalization constant of the $i$th
incoming partonic field external to $h_I$, and $Z_{S,LI}$ is
a matrix of renormalization constants, which describe the
renormalization of the soft function, including
mixing of color structures.  $Z_{S}$ is defined to include the
wave function renormalization
necessary for the outgoing eikonal lines that
represent the heavy quarks.

From Eq.\ (\ref{HSren}), we see that 
the soft function  $S_{LI}$ satisfies the
renormalization group equation~\cite{KS}
\begin{equation}
\left(\mu {\partial \over \partial \mu}+\beta(g){\partial \over \partial g}
\right)\,S_{LI}
=-(\Gamma^\dagger_S)_{LB}S_{BI}-S_{LA}(\Gamma_S)_{AI}\, ,
\label{RGE}
\end{equation}
where $\Gamma_S$ is an anomalous dimension matrix 
that is calculated by explicit renormalization of the soft function.
In a minimal subtraction renormalization scheme and with
$\epsilon=4-n$, where $n$ is the number of space-time dimensions,
the matrix of anomalous dimensions at one loop is given by
\begin{equation}
\Gamma_S (g)=-\frac{g}{2} \frac {\partial}{\partial g}{\rm Res}_{\epsilon
\rightarrow 0} Z_S (g, \epsilon) \, .
\end{equation}
Explicit results for the soft anomalous dimension
matrices $\Gamma_S$ for the partonic subprocesses involved in 
heavy quark production will be presented in the next two
sections.

The renormalization group equation (\ref{RGE})
is, in general, a matrix equation. Its solution is
\beqa
&& \hspace{-15mm}  
{\rm Tr}\left\{H\left({Q\over\mu},\theta,\alpha_s(\mu^2)
\right) \;
{\tilde S} \left({Q\over N\mu},\theta,\alpha_s(\mu^2) \right)
\right\}
\nonumber \\ && \hspace{-12mm}
= {\rm Tr}\left\{ 
H\left({Q\over\mu},\theta,\alpha_s(\mu^2)\right) \;
\bar{P} \exp \left[\int_\mu^{Q/N} {d\mu' \over \mu'}\; 
\Gamma_S{}^\dagger\left(\alpha_s(\mu'^2)\right)\right] \right.
\nonumber\\ && \hspace{-12mm} \left. \times \;
{\tilde S} \left(1,\theta,\alpha_s\left(Q^2/N^2\right)\right)\;
P \exp \left[\int_\mu^{Q/N} {d\mu' \over \mu'}\; \Gamma_S
\left(\alpha_s(\mu'^2)\right)\right]
\right\} \, ,
\label{rgesol}
\eeqa
where the trace is taken in color space, so that
\beq
{\rm Tr}[H {\tilde S}]=H_{IL}{\tilde S}_{LI}=h_I h^*_L {\tilde S}_{LI} \, .
\eeq
At lowest order, ${\tilde S}_{LI}={\rm Tr} [c_L^{\dagger} c_I]$.
The symbols $P$ and $\bar{P}$ refer
to path ordering in the same sense as the integration variable $\mu'$ and 
against it, respectively.

\subsubsection{Resummed cross section}

Using Eqs.~(\ref{psiphiHS}), (\ref{psiphimu}), and (\ref{rgesol}), 
the resummed heavy quark cross section in moment space is then 
\beqa
\tilde{{\sigma}}_{f{\bar f}\rightarrow Q{\bar Q}}(N) &=&  
R_{(f)}^2(\alpha_s(\mu^2)) \; 
\exp \left \{2 \left[ E^{(f_i)}(N,Q^2) \right. \right. 
\nonumber\\&& \hspace{15mm} \left. \left.
{}-2\int_\mu^Q{d\mu'\over\mu'}\; 
\left[\gamma_{f_i}(\alpha_s(\mu'{}^2))-\gamma_{f_if_i}(N,\alpha_s(\mu'{}^2)) 
\right] \right] \right\}
\nonumber\\ && \times \; {\rm Tr} \left \{
H\left({Q\over\mu},\theta,\alpha_s(\mu^2)\right) \;
\bar{P} \exp \left[\int_\mu^{Q/N} {d\mu' \over \mu'} \;
\Gamma_S^\dagger\left(\alpha_s(\mu'^2)\right)\right] \right.
\nonumber\\ && \times \; \left.
{\tilde S} \left(1,\theta,\alpha_s(Q^2/N^2) \right) \; 
P \exp \left[\int_\mu^{Q/N} {d\mu' \over \mu'}\; \Gamma_S
\left(\alpha_s(\mu'^2)\right)\right] \right\}\, .
\nonumber \\
\label{resHQ}
\eeqa

At the level of leading logarithms of $N$ in the soft gluon function
$S$, and therefore at next-to-leading logarithm of $N$ in
the cross section as a whole, we may simplify this result by
choosing a color basis in which the anomalous dimension matrix 
$\Gamma_S$ is diagonal, 
with eigenvalues $\lambda_I$ for each basis color
tensor labelled by $I$.
Then, we have
\begin{eqnarray}
{\tilde S}_{LI}\left(\frac{Q}{N\mu}, \theta, \alpha_s(\mu^2)\right)&=&
{\tilde S}_{LI}\left(1,\theta, \alpha_s\left(\frac{Q^2}{N^2}\right)\right)
\nonumber \\ &&  \hspace{-15mm} \times \, 
\exp\left[-\int^{\mu}_{Q/N}\frac{d \bar{\mu}}{\bar{\mu}}
[\lambda_I(\alpha_s(\bar{\mu}^2))
+\lambda^*_L(\alpha_s(\bar{\mu}^2))]\right]\, .
\label{rgedgsol}
\end{eqnarray}

Thus, in a diagonal basis, and with $\mu=Q$ and $R_{(f)}$ normalized to unity,
we can rewrite the resummed cross section in a simplified form as 
\beqa
\hspace{-10mm}\tilde{{\sigma}}_{f{\bar f}\rightarrow Q{\bar Q}}(N) &=& 
H_{IL}\left({Q\over\mu},\theta,\alpha_s(\mu^2)\right) \;
{\tilde S}_{LI} \left(1,\theta,\alpha_s(Q^2/N^2) \right)
\nonumber \\ && \times \, 
\exp \left[E_{LI}^{(f {\bar f})}(N,\theta,Q^2)\right] \, ,
\label{crossdiag}
\eeqa
where the exponent is
\begin{eqnarray}
E_{LI}^{(f {\bar f})}(N,\theta,Q^2)&=&-\int_0^1 dz \frac{z^{N-1}-1}{1-z}
\left\{\int^{(1-z)^{m_S}}_{(1-z)^2} \frac{d\lambda}{\lambda}
g_1^{(f{\bar f})}[\alpha_s(\lambda Q^2)] \right.
\nonumber \\ &&  \hspace{-3mm}\left.
{}+g_2^{(f{\bar f})}[\alpha_s((1-z)^{m_S} Q^2)]
+g_3^{(IL,f{\bar f})}[\alpha_s((1-z)^2 Q^2),\theta] \right\}\, ,
\nonumber \\
\label{ELI}
\end{eqnarray}
with the functions $g_1$, $g_2$, and $g_3$ defined by
\beqa
g_1^{(f\bar f)}=A^{(f)}+A^{(\bar f)} \, , 
\quad g_2^{(f \bar f)}=B^{(f)}+B^{(\bar f)} \, ,
\nonumber \\
\quad g_3^{(IL,f {\bar f})}=-\lambda_I-\lambda_L^*+\frac{1}{2}\nu^{(f)}
+\frac{1}{2}\nu^{(\bar f)} \, .
\label{g1g2g3}
\eeqa

In the next two sections we present the soft anomalous dimension matrices
for heavy quark production through light quark annihilation and gluon fusion,
and we give one- and two-loop expansions of the resummed cross section.

\mysection{Soft anomalous dimension matrix for 
\protect\newline the process $q \bar{q} \rightarrow Q \bar{Q}$}

\subsection{Eikonal diagrams and $\Gamma_S$ 
for $q \bar{q} \rightarrow Q \bar{Q}$}

We begin with the soft anomalous dimension matrix for heavy quark
production through light quark annihilation,
\begin{equation}
q(p_a,r_a)+{\bar q}(p_b,r_b) \rightarrow {\bar Q}(p_1,r_1) + Q(p_2,r_2)\, ,
\end{equation}
where the $p_i$'s and $r_i$'s denote momenta and colors of the partons
in the process.
The anomalous dimension matrix for this process is $2 \times 2$,
since the color indices $I$ and $L$ range over two values.
We introduce the Mandelstam invariants
\begin{equation}
s=(p_a+p_b)^2\, , \quad t_1=(p_a-p_1)^2-m^2\, , \quad u_1=(p_b-p_1)^2-m^2\, ,
\label{Mandelstam}
\end{equation}
with $m$ the heavy quark mass, which satisfy
$s+t_1+u_1=0$ at partonic threshold.
The UV divergent contribution to $S_{LI}$ is the sum of graphs in Figs. 2(a)
and 2(b). We give details of the calculation in the Appendix.

\begin{figure}
\centerline{
\psfig{file=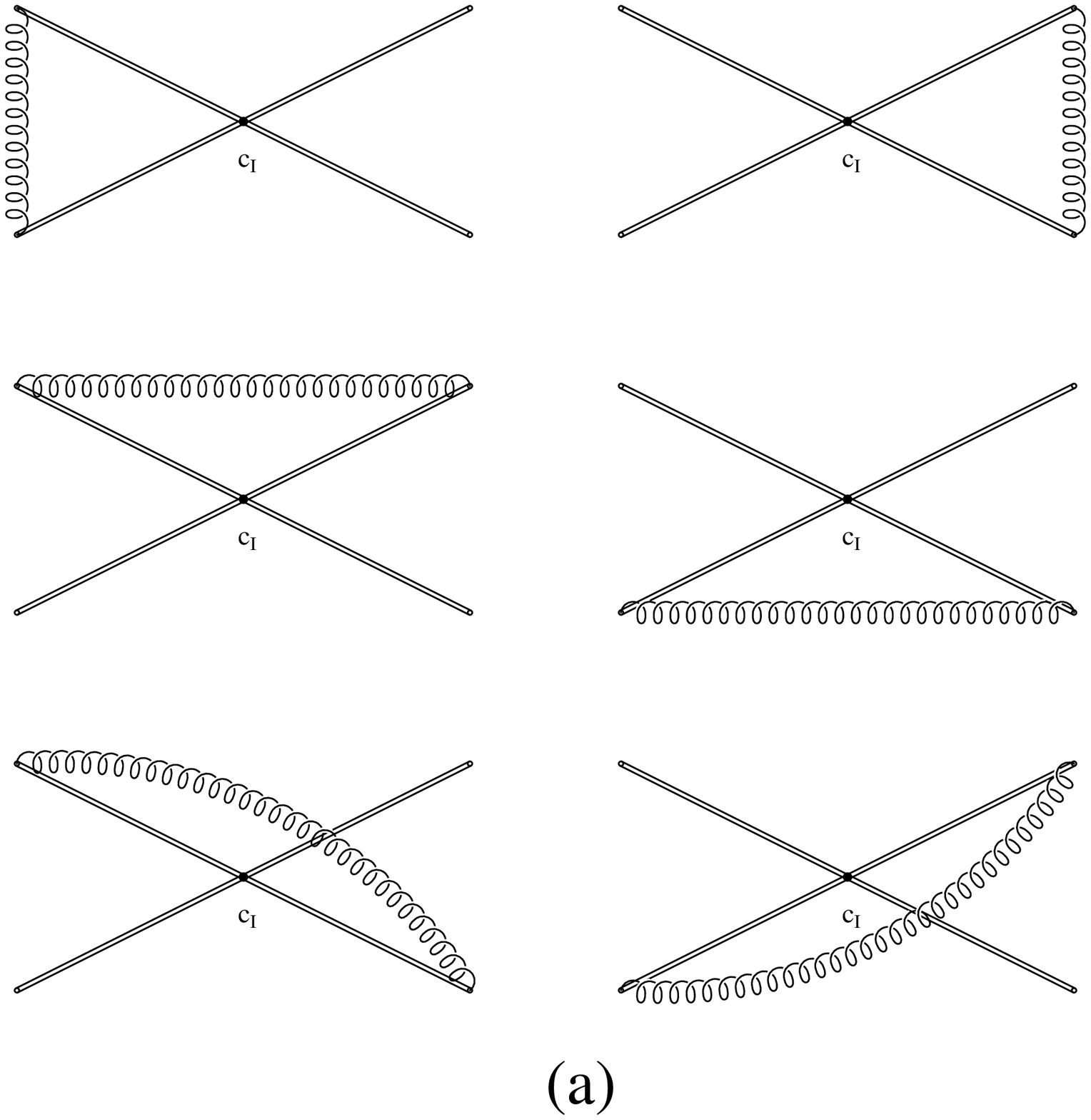,height=4.05in,width=4.05in,clip=}}
\vspace{13mm}
\centerline{
\psfig{file=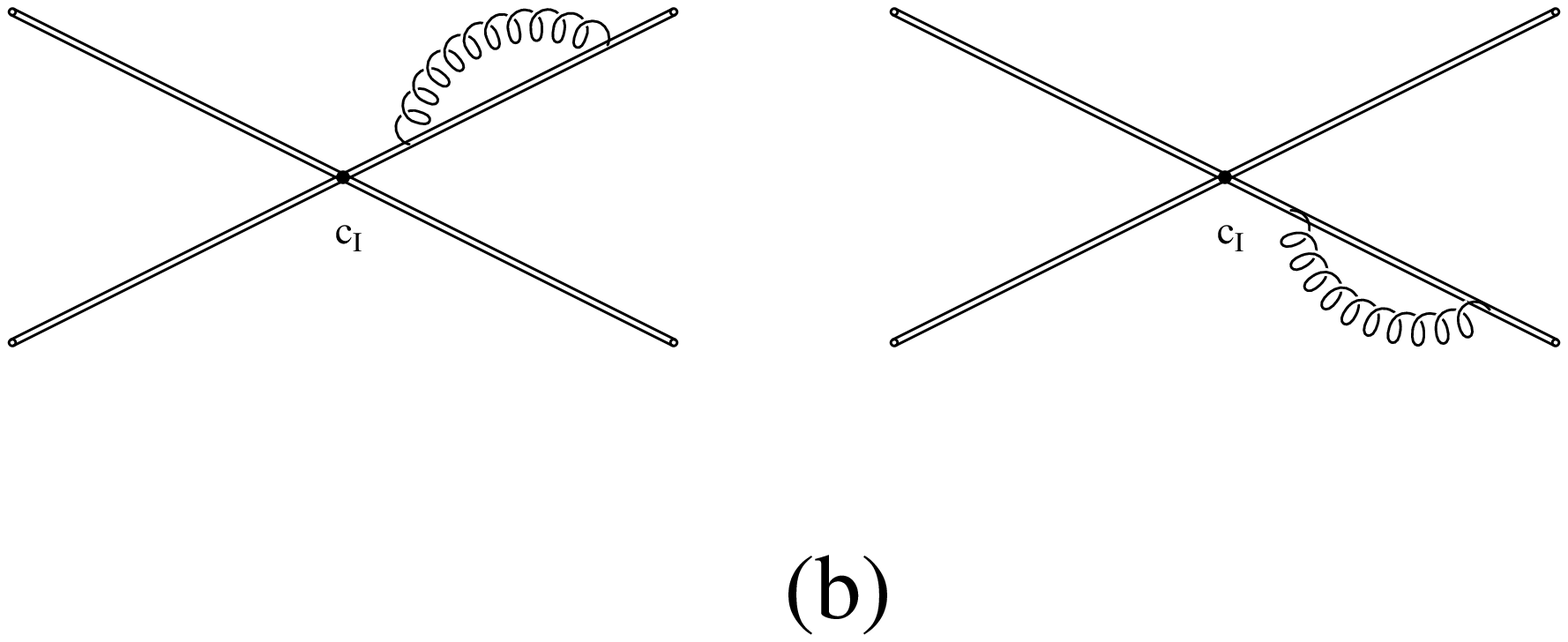,height=1.5in,width=4.05in,clip=}}
{Fig. 2. One-loop corrections to $S_{LI}$ for partonic subprocesses
in heavy quark or dijet production: 
(a) eikonal vertex corrections;
(b) eikonal self-energy graphs for heavy quark production.}
\label{fig2}
\end{figure}
 
\begin{figure}
\centerline{\hspace{13mm}
\psfig{file=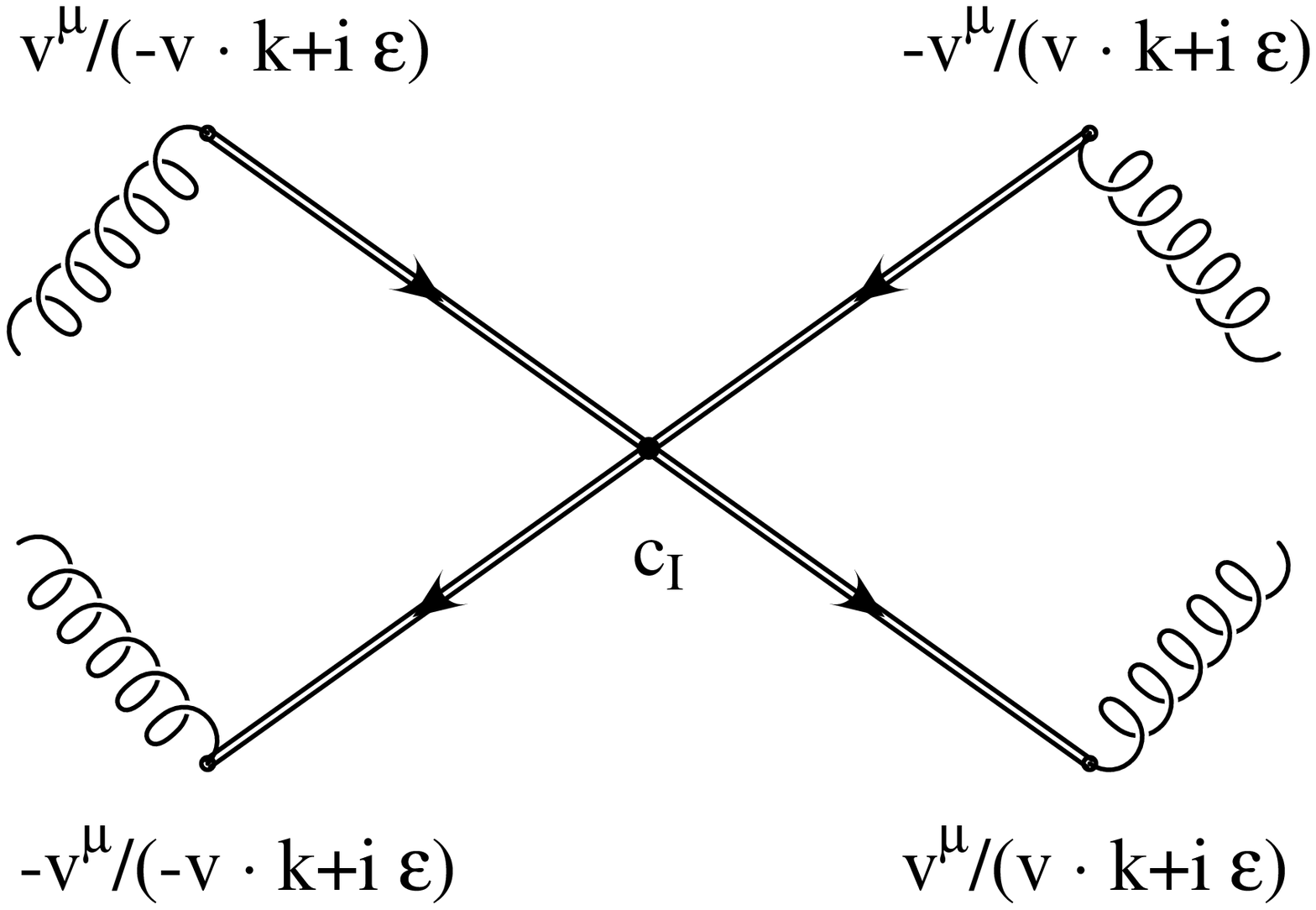,height=2.5in,width=4.05in,clip=}}
{Fig. 3. Feynman rules for eikonal lines 
representing quarks, in the process $q{\bar q} \rightarrow Q{\bar Q}$. 
The gluon momentum flows out of the eikonal lines.
Group matrices at the vertices are the same as
for quark lines.}
\label{fig3}
\end{figure}

In our calculations we use Feynman rules for eikonal diagrams
in axial gauge as shown in Fig. 3 (resummation can be performed in
a covariant gauge as well~\cite{Li}).
We define dimensionless vectors $v_i^{\mu}$  by
$p_i^{\mu}=Qv_i^{\mu}/\sqrt{2}$,
which obey $v_i^2=0$ for the light incoming quarks and
$v_i^2=2m^2/Q^2$ for the outgoing heavy quarks.
The propagator for a quark, antiquark or gluon eikonal line 
is then given by
\beq
\frac{i}{\delta v \cdot q + i\epsilon} \, ,
\label{qprop}
\eeq
where $\delta=+1(-1)$ for the momentum $q$ flowing 
in the same (opposite) direction as the dimensionless vector $v$.
The interaction vertex for a quark or antiquark eikonal line is 
\beq
-ig_s \left({T_F^c}\right)_{ba} v^{\mu} \Delta \, ,
\eeq 
with $\Delta=+1(-1)$ for a quark (antiquark).
The $T_F^c$ are the generators of $SU(3)$ in the fundamental representation.

For the determination of $\Gamma_S$ an appropriate choice of color basis 
has to be made, e.g. singlet exchange in the $s$- and $u$-channels,
or singlet and octet exchange in the $s$-channel. 
Here we give the result in a general axial gauge 
in the $s$-channel singlet-octet basis:
\beqa
c_1 &=& c_{\rm singlet}=\delta_{r_ar_b}\delta_{r_1r_2}\, , \quad \quad
\nonumber \\ 
c_2 &=& c_{\rm octet}=(T^c_F)_{r_b r_a}(T^c_F)_{r_2 r_1}
=-\frac{1}{2N}c_{\rm singlet}
+\frac{1}{2} \delta_{r_ar_2}\delta_{r_br_1}\, .
\label{18basis}
\eeqa
The results of our calculation are~\cite{Thesis,NKGS,KS} 
\begin{equation}
\Gamma_S=\left[\begin{array}{cc}
\Gamma_{11} & \Gamma_{12} \\
\Gamma_{21} & \Gamma_{22}
\end{array}
\right] \, ,
\label{matrixqqQQ}
\end{equation}
with 
\begin{eqnarray}
\Gamma_{11}&=&-\frac{\alpha_s}{\pi}C_F \, [L_{\beta}
+\ln(2\sqrt{\nu_a\nu_b})+\pi i] \, ,
\nonumber \\
\Gamma_{21}&=&\frac{2\alpha_s}{\pi}
\ln\left(\frac{u_1}{t_1}\right) \, ,
\nonumber \\
\Gamma_{12}&=&\frac{\alpha_s}{\pi}
\frac{C_F}{C_A} \ln\left(\frac{u_1}{t_1}\right) \, ,
\nonumber \\
\Gamma_{22}&=&\frac{\alpha_s}{\pi}\left\{C_F
\left[4\ln\left(\frac{u_1}{t_1}\right)
-\ln(2\sqrt{\nu_a\nu_b})
-L_{\beta}-\pi i\right]\right.
\nonumber \\ &&
\left.{}+\frac{C_A}{2}\left[-3\ln\left(\frac{u_1}{t_1}\right)
-\ln\left(\frac{m^2s}{t_1u_1}\right)+L_{\beta}+\pi i \right]\right\}\, ,
\label{GammaqqQQ}
\end{eqnarray}
where $L_\beta$ is the  velocity-dependent
eikonal function
\begin{equation}
L_{\beta}=\frac{1-2m^2/s}{\beta}\left(\ln\frac{1-\beta}{1+\beta}
+\pi i \right)\, ,
\end{equation}
with $\beta=\sqrt{1-4m^2/s}$. 
Note that here we haven't absorbed the function $\nu^{(f)}$, Eq.~(\ref{nuf}), 
in the results for $\Gamma_S$, as was done in Refs.~\cite{Thesis,NKGS,KS}.
The gauge dependence of the incoming eikonal lines is given in terms of   
\beq
\nu_i \equiv \frac{(v_i \cdot n)^2}{|n|^2} \, .
\label{nui}
\eeq
This gauge dependence cancels against corresponding terms in the
parton distributions.
The gauge dependence of the outgoing heavy quarks is cancelled
by the inclusion of the self-energy diagrams in Fig. 2(b) (see also
the discussion in the Appendix). 

$\Gamma_S$ is diagonalized in this singlet-octet basis
at absolute threshold, $\beta=0$, and also for arbitrary $\beta$ 
when the parton-parton c.m. scattering angle is
$\theta=90^\circ$ (where $u_1=t_1$), with
eigenvalues that may be read off from Eq.\ (\ref{GammaqqQQ}).

In general, the eigenvalues of the anomalous dimension matrix,
Eq.~(\ref{matrixqqQQ}), are
\begin{equation}
\lambda_{1,2}=\frac{1}{2}\left[\Gamma_{11}+\Gamma_{22}
\pm \sqrt{(\Gamma_{11}-\Gamma_{22})^2+4 \Gamma_{12} \Gamma_{21}}\right] \, ,
\label{qqQQev}
\end{equation}
and the eigenvectors are
\begin{equation}
e_i=\left[\begin{array}{c}
\frac{\Gamma_{12}}{\lambda_i-\Gamma_{11}} \vspace{2mm}\\
1
\label{qqQQevec}
\end{array}\right]
\end{equation}
for each eigenvalue $\lambda_i$.
These expressions will be useful when we discuss the diagonalization
of the anomalous dimension matrices in Section 5.

\subsection{One- and two-loop expansions of the resummed 
\protect\newline cross section for $q{\bar q} \rightarrow Q {\bar Q}$}

We can expand the resummed heavy quark cross section to any fixed order
in perturbation theory without having to diagonalize the soft anomalous
dimension matrices. 
Such fixed-order expansions of resummed cross sections have also
been done at one and two loops for the Drell-Yan process~\cite{Magnea}, 
Higgs production~\cite{Higgsres}, 
and heavy quark electroproduction~\cite{LM}.
We will present one- and two-loop expansions of the resummed
cross sections for direct photon and $W$ boson production in Section 10.

First we give the one-loop expansion for 
$q {\bar q} \rightarrow Q {\bar Q}$. At one-loop
the inverse Mellin transforms are trivial. 
The result is proportional to the Born cross section, 
$\sigma^B_{q{\bar q}\rightarrow Q{\bar Q}}$,
and the one-loop contributions from $g_1^{(q{\bar q})}$
and $g_2^{(q{\bar q})}$, Eq.~(\ref{g1g2g3}),
and ${\rm Re} \, \Gamma_{22}$, the real part of $\Gamma_{22}$.
In the DIS scheme the one-loop result is
\begin{eqnarray}
{\hat \sigma}^{\rm DIS \, (1)}_{q{\bar q}\rightarrow Q{\bar Q}}
(1-z,m^2,s,t_1,u_1)&=&\sigma^B_{q{\bar q}\rightarrow Q{\bar Q}}
\frac{\alpha_s}{\pi}\left\{2C_F\left[\frac{\ln(1-z)}{1-z}\right]_{+}
\right.
\nonumber \\ && \hspace{-30mm}
{}+\left[\frac{1}{1-z}\right]_{+} \left[C_F\left(\frac{3}{2} 
+8\ln\left(\frac{u_1}{t_1}\right)
-2-2 \, {\rm Re} \, L_{\beta}+2\ln\left(\frac{s}{\mu^2}\right)\right)\right.
\nonumber \\ && \hspace{-20mm}
\left.\left.
{}+C_A\left(-3\ln\left(\frac{u_1}{t_1}\right)+{\rm Re} \, L_{\beta}
-\ln\left(\frac{m^2s}{t_1 u_1}\right)\right)\right]\right\}\, ,
\end{eqnarray}
while in the $\overline{\rm MS}$ scheme
\begin{eqnarray}
{\hat \sigma}^{\overline{\rm MS} \, (1)}_{q{\bar q}\rightarrow Q{\bar Q}}
(1-z,m^2,s,t_1,u_1)&=&\sigma^B_{q{\bar q}\rightarrow Q{\bar Q}}
\frac{\alpha_s}{\pi}\left\{4C_F\left[\frac{\ln(1-z)}{1-z}\right]_{+}
\right.
\nonumber \\ && \hspace{-30mm}
{}+\left[\frac{1}{1-z}\right]_{+} 
\left[C_F\left(8\ln\left(\frac{u_1}{t_1}\right)
-2-2 \, {\rm Re} \, L_{\beta}+2\ln\left(\frac{s}{\mu^2}\right)\right)\right.
\nonumber \\ && \hspace{-20mm}
\left.\left.
{}+C_A\left(-3\ln\left(\frac{u_1}{t_1}\right)+{\rm Re} \, L_{\beta}
-\ln\left(\frac{m^2s}{t_1 u_1}\right)\right)\right]\right\}\, .
\end{eqnarray}

Note that these expressions for the cross section are for fixed values of
the $Q{\bar Q}$ invariant mass.
There are approximate one-loop results for heavy quark production
in the literature for a single-particle
inclusive cross section where the singular behavior is given in terms
of the variable $s_4=s+t_1+u_1$ \cite{mengetal} .
Therefore there are small differences in the results for the two cross
sections due to the different phase spaces.
Nevertheless, the cross sections are kinematically equivalent in the
limit $\beta\rightarrow 0$, where $s_4=2m^2(1-z)$.

If one uses single-particle inclusive kinematics in the resummation
procedure \cite{LOS}, then one reproduces the results in 
Ref. \cite{mengetal} exactly. In addition one can add the Coulomb 
corrections to the one-loop expansions~\cite{pp,KLMV}
(these corrections also appear in Ref. \cite{mengetal}).
Numerically the one-loop expansions of the NLL resummed cross section
are very good approximations to the exact NLO cross sections at
both the partonic and hadronic levels \cite{NKRV, KLMV}.

One can expand the resummed cross section to higher orders and thus
make predictions of perturbation theory at these higher orders near threshold. 
The inversion back to momentum space at any fixed order is straightforward.
For the two-loop expansion of the resummed cross section
in the $\overline {\rm MS}$ scheme and with $\mu=Q$ we have 
\begin{eqnarray}
{\hat \sigma}^{\overline {\rm MS} \, (2)}_{q{\bar q}\rightarrow Q{\bar Q}}
(1-z,m^2,s,t_1,u_1)&=&
\sigma^B_{q{\bar q}\rightarrow Q{\bar Q}} 
\frac{\alpha_s^2}{\pi^2} 
\left\{8 C_F^2 \left[\frac{\ln^3(1-z)}{1-z}\right]_{+} \right.
\nonumber \\ && \hspace{-40mm}
{}+\left[\frac{\ln^2(1-z)}{1-z}\right]_{+} C_F \left[-\beta_0 
+12C_F \left(4 \ln\left(\frac{u_1}{t_1}\right)-{\rm Re}\, 
L_{\beta}-1\right) \right.
\nonumber \\ && \hspace{-25mm} \left. \left.
{}+6 C_A \left(-3 \ln\left(\frac{u_1}{t_1}\right) 
-\ln\left(\frac{m^2s}{t_1 u_1}\right) 
+{\rm Re} \, L_{\beta}\right)\right]\right\} \, .
\end{eqnarray}
Analogous results have been obtained in the DIS scheme, and also
in single-particle inclusive kinematics in both schemes~\cite{KLMV}. 

\mysection{Soft anomalous dimension matrix for 
\protect\newline the process $gg \rightarrow Q {\bar Q}$}

\subsection{Eikonal diagrams and $\Gamma_S$ 
for $gg \rightarrow Q \bar{Q}$}

We continue with the soft anomalous dimension matrix for heavy quark
production through gluon fusion,
\begin{equation}
g(p_a,r_a)+g(p_b,r_b) \rightarrow {\bar Q}(p_1,r_1) + Q(p_2,r_2)\, ,
\end{equation}
with momenta and colors labelled by the $p_i$'s and $r_i$'s, respectively. 
Here $\Gamma_S$ is a $3 \times 3$ matrix.
We use the same Mandelstam invariants as in the previous section,
Eq.~(\ref{Mandelstam}).
The UV divergent graphs are the same as in Fig.~2, where now the incoming
eikonal lines represent gluons.
In our calculations we use the eikonal rules as shown in Fig. 4
(see also the previous section). The gluon eikonal vertex is
\beq
-g_s f^{abc} v^{\mu} \Delta \, ,
\eeq
where we read the color indices $a,b,c$ anticlockwise, and where 
$\Delta=+1(-1)$ for the gluon located below (above) the eikonal line.

\begin{figure}
\centerline{\hspace{13mm}
\psfig{file=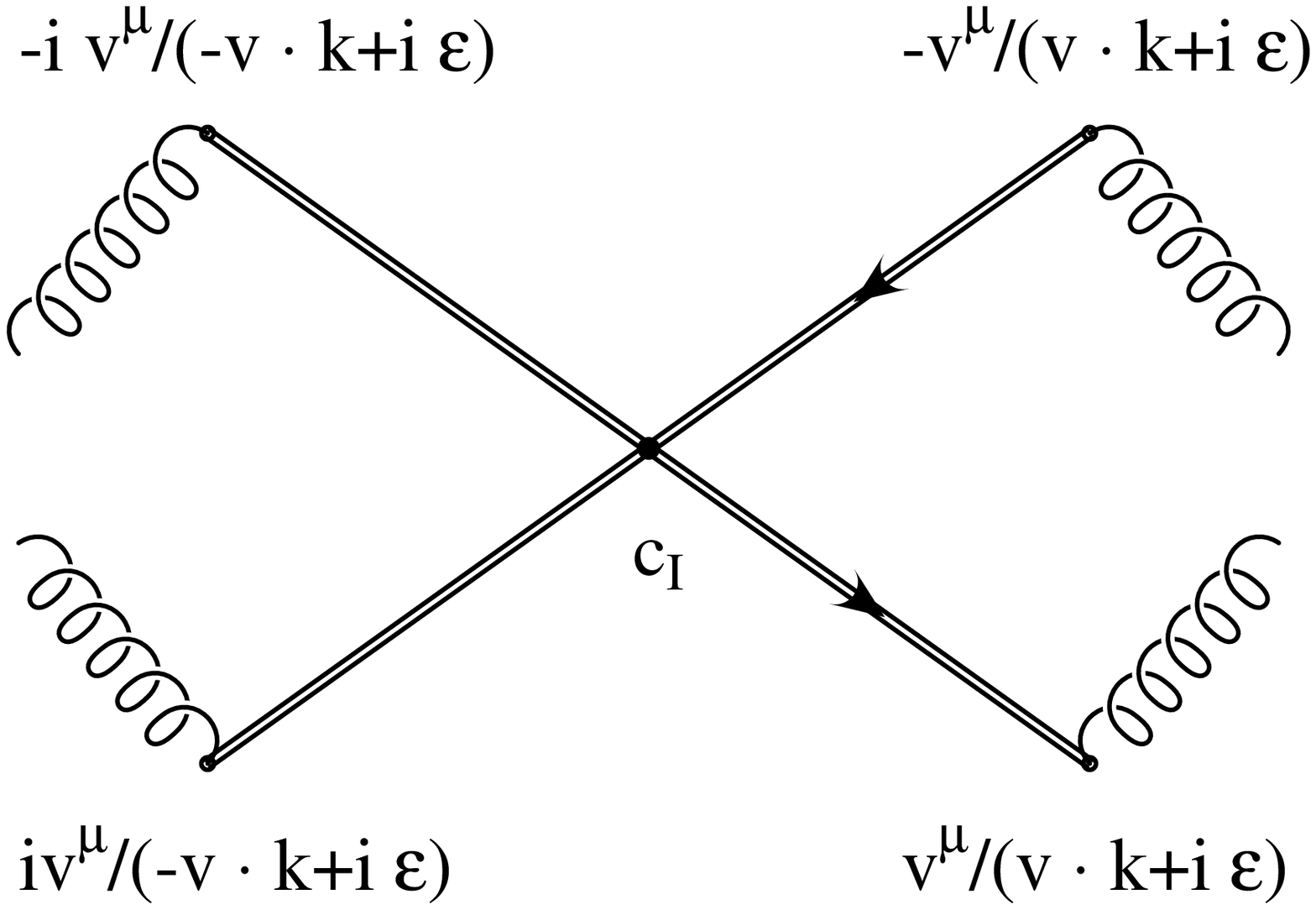,height=2.5in,width=4.05in,clip=}}
{Fig. 4. Feynman rules for eikonal lines representing gluons, in the
process $gg \rightarrow Q{\bar Q}$.
The gluon momentum flows out of the eikonal lines. Group
matrices at the vertices on the left side are those of three-gluon 
vertices; those on the right are as for quark lines.}
\label{fig4}
\end{figure}

We make the following choice for the color basis:
\begin{equation}
c_1=\delta_{r_ar_b}\,\delta_{r_2r_1}, \quad 
c_2=d^{r_ar_bc}\,(T^c_F)_{r_2r_1},
\quad c_3=i f^{r_ar_bc}\,(T^c_F)_{r_2r_1} \, ,
\end{equation}
where again the $T^c_F$ are the generators of $SU(3)$ in the fundamental
representation, and $d^{abc}$ and $f^{abc}$ are the totally
symmetric and antisymmetric $SU(3)$ invariant tensors, respectively.
The anomalous dimension matrix in this color basis and in a 
general axial gauge is given by~\cite{Thesis,KS}
\begin{equation}
\Gamma_S=\left[\begin{array}{ccc}
\Gamma_{11} & 0 & \frac{\Gamma_{31}}{2} \vspace{2mm} \\
0 & \Gamma_{22} & \frac{N_c}{4} \Gamma_{31} \vspace{2mm} \\
\Gamma_{31} & \frac{N_c^2-4}{4N_c}\Gamma_{31} & \Gamma_{22}
\end{array}
\right] \, ,
\label{GammaggQQ33}
\end{equation}
where
\begin{eqnarray}
\Gamma_{11}&=&\frac{\alpha_s}{\pi}\left [-C_F(L_{\beta}+1)
+C_A\left(-\frac{1}{2}\ln\left({4\nu_a\nu_b}\right)+1-\pi i\right)\right ],
\nonumber \\
\Gamma_{31}&=&\frac{\alpha_s}{\pi}\ln\left(\frac{u_1^2}{t_1^2}\right) \, ,
\nonumber \\
\Gamma_{22}&=&\frac{\alpha_s}{\pi}\left\{-C_F(L_{\beta}+1)
+\frac{C_A}{2}\left[-\ln\left(4\nu_a\nu_b\right)
+2+\ln\left(\frac{t_1 u_1}{m^2 s}\right)+L_{\beta}-\pi i
\right]\right\}.
\nonumber \\
\label{GammaggQQ}
\end{eqnarray}
Here we haven't absorbed the function 
$\nu^{(f)}$, Eq.~(\ref{nuf}), in the results for $\Gamma_S$, 
as was done in Refs.~\cite{Thesis,KS}.
Again we note that $\Gamma_S$ is diagonalized in this basis
at absolute threshold, $\beta=0$, and also for arbitrary $\beta$, 
when the parton-parton c.m. scattering angle is
$\theta=90^\circ$ , with
eigenvalues that may be read off from Eq.\ (\ref{GammaggQQ}).

The eigenvalues of $\Gamma_S$ may be written as
\begin{eqnarray}
\lambda_1&=&\frac{1}{3}
(X^{1/3}-Y+\Gamma_{11}+2\Gamma_{22}) ,
\nonumber \\
\lambda_{2,3}&=&\frac{1}{3}\left[-\frac{1}{2}(X^{1/3}-Y)+\Gamma_{11}
+2\Gamma_{22} \pm \frac{1}{2} i \sqrt{3} (X^{1/3}+Y)\right] ,
\label{ggQQev}
\end{eqnarray}
where
\begin{eqnarray}
X&=&(\Gamma_{11}-\Gamma_{22})^3
-\frac{9}{16}(N_c^2-8)\Gamma_{31}^2(\Gamma_{11}-\Gamma_{22})
\nonumber \\ &&
{}+\frac{3\sqrt{3}}{4}\left[
-\frac{(N_c^2+4)^3}{256}\Gamma_{31}^6
-(N_c^2-4)\Gamma_{31}^2(\Gamma_{11}-\Gamma_{22})^4 \right.
\nonumber \\ && \left.
{}+\frac{1}{8}((N_c^2-8)^2-12(N_c^2-2)) \Gamma_{31}^4
(\Gamma_{11}-\Gamma_{22})^2
\right]^{1/2}
\end{eqnarray}
and
\begin{equation}
Y=-\left[(\Gamma_{11}-\Gamma_{22})^2
+\frac{3}{16}\Gamma_{31}^2(N_c^2+4)\right]X^{-1/3} .
\end{equation}

The eigenvectors of $\Gamma_S$ are given by
\begin{equation}
e_i=\left[\begin{array}{c}
\frac{\Gamma_{31}}{2(\lambda_i-\Gamma_{11})} \vspace{2mm} \\
\frac{N_c\Gamma_{31}}{4(\lambda_i-\Gamma_{22})} \vspace{2mm} \\
1
\end{array}\right] \, ,
\label{ggQQevec}
\end{equation}
for each eigenvalue $\lambda_i$.

\subsection{One- and two-loop expansions of the resummed 
\protect\newline cross section for $gg \rightarrow Q {\bar Q}$}

Again, we may expand the resummed cross section for 
$gg \rightarrow Q {\bar Q}$ at one and two loops or higher.
In this case the color decomposition is more complicated~\cite{Thesis,KS}.
The one-loop expansion in the $\overline {\rm MS}$ scheme is
\beqa
{\hat \sigma}^{\overline {\rm MS} \, (1)}_{gg \rightarrow Q{\bar Q}}
(1-z,m^2,s,t_1,u_1)&=&
\sigma^B_{gg\rightarrow Q{\bar Q}} \frac{\alpha_s}{\pi} 
\left\{4C_A\left[\frac{\ln(1-z)}{1-z}\right]_{+}\right.
\nonumber \\ && \hspace{25mm} \left.
{}-2C_A \ln\left(\frac{\mu^2}{s}\right) 
\left[\frac{1}{1-z}\right]_{+}\right\}
\nonumber \\ && \hspace{-50mm}
{}+\alpha_s^3 K_{gg} B_{QED} \left[\frac{1}{1-z}\right]_{+}
\left\{N_c(N_c^2-1)\frac{(t_1^2+u_1^2)}{s^2}
\left[\left(-C_F+\frac{C_A}{2}\right)
{\rm Re} \, L_{\beta}\right. \right.
\nonumber \\ && \hspace{-35mm} \left.
{}+\frac{C_A}{2}\ln\left(\frac{t_1u_1}{m^2s}\right)
-C_F\right]+\frac{N_c^2-1}{N_c}(C_F-C_A) {\rm Re} \, L_{\beta}
\nonumber \\ && \hspace{-50mm} \left.
{}-(N_c^2-1)\ln\left(\frac{t_1u_1}{m^2s}\right)
+C_F \frac{N_c^2-1}{N_c}
+\frac{N_c^2}{2}(N_c^2-1)
\ln\left(\frac{u_1}{t_1}\right)\frac{(t_1^2-u_1^2)}{s^2} \right\} \, ,
\nonumber\\&&
\label{gg1loop}
\eeqa
where $\sigma^B_{gg\rightarrow Q{\bar Q}}$ is the Born cross section,
\beq
B_{\rm QED}=\frac{t_1}{u_1}+\frac{u_1}{t_1}+\frac{4m^2s}{t_1u_1}
\left(1-\frac{m^2s}{t_1u_1}\right) \, ,
\eeq
and $K_{gg}=(N^2-1)^{-2}$ is a color average factor.

As we explained in the previous section,
we cannot compare our one-loop expansion directly with the approximate
NLO results of Ref. \cite{mengetal}, but as $\beta\rightarrow 0$ our 
expression becomes identical to the $\beta\rightarrow 0$ limit of the
results in \cite{mengetal}. Again, we note that even for $\beta>0$
our one-loop expansion is nearly the same as in Ref.~\cite{mengetal}. 
If one uses single-particle inclusive kinematics in the resummation
procedure then the agreement with Ref.~\cite{mengetal} is exact~\cite{KLMV}.

The two-loop expansion of the resummed cross section 
in the $\overline {\rm MS}$ scheme and with $\mu=Q$ is
\beqa
{\hat \sigma}^{\overline {\rm MS} \, (2)}_{gg \rightarrow Q{\bar Q}}
(1-z,m^2,s,t_1,u_1)&=&
\sigma^B_{gg\rightarrow Q{\bar Q}} \frac{\alpha_s^2}{\pi^2}
\left\{8C_A^2\left[\frac{\ln^3(1-z)}{1-z}\right]_{+}\right.
\nonumber \\ && \quad \quad \quad \left.
-\beta_0 C_A \left[\frac{\ln^2(1-z)}{1-z}\right]_{+}\right\}
\nonumber \\ && \hspace{-55mm}
{}+\frac{\alpha_s^4}{\pi} K_{gg} B_{\rm QED} 
\left[\frac{\ln^2(1-z)}{1-z}\right]_{+} C_A \, 3(N_c^2-1)
\left\{\frac{(t_1^2+u_1^2)}{s^2} \right.
\nonumber \\ && \hspace{-55mm} \quad \times \, 
\left[N_c^2 \ln\left(\frac{t_1u_1}{m^2s}\right)
-2N_c\left(C_F-\frac{C_A}{2}\right){\rm Re}L_{\beta}-2 N_c C_F\right]
+2\frac{C_F}{N_c}
\nonumber \\ && \hspace{-60mm} \quad \left.
{}+2\ln\left(\frac{m^4}{t_1 \, u_1}\right)
+2\frac{1}{N_c}(C_F-C_A) \,{\rm Re} \, L_{\beta}
+N_c^2\frac{(t_1^2-u_1^2)}{s^2}\ln\left(\frac{u_1}{t_1}\right)\right\}. 
\eeqa
Analogous results may also be obtained in single-particle inclusive
kinematics~\cite{KLMV}.

\mysection{Diagonalization procedure and 
\protect\newline numerical results}

As we saw in the previous two sections, the soft anomalous
dimension matrices, $\Gamma_S$,  are in general not diagonal. 
They are only diagonal at absolute threshold, $\beta \rightarrow 0$, 
and at a scattering angle $\theta=90^{\circ}$. In these cases the
exponentiated cross section has a simpler form. Numerical studies of
the resummed cross section at $\theta=90^{\circ}$ for top quark 
production at the Fermilab Tevatron and bottom quark production
at HERA-B were presented in Ref.~\cite{NKJSRV}.

In general, however, we must find new color bases where
$\Gamma_S$ is diagonal so that the resummed cross section 
can take the  simpler form of Eq.~(\ref{crossdiag}).
In this section we outline the required diagonalization procedure 
and apply it to heavy quark production, giving some numerical 
results for top quark production at the Tevatron.

\subsection{General diagonalization procedure}

The Born cross section can be written in an arbitrary color basis as
a product of the hard components and the color tensors:
\beq
\sigma^B=H_{IJ} \, c_J^{\dagger} \, c_I \, .
\eeq
In a diagonal basis, $c_I'=c_K R_{KI}$,  
the Born cross section can be rewritten as
\beq
\sigma^B=H_{IJ} \, (R^{-1})^{\dagger}_{JL} (c')^{\dagger}_L
c_K' (R^{-1})_{KI}  \, ,
\eeq
where $R$ is made from the eigenvectors of the anomalous dimension matrix,
and we have used $c_I=c_K' R^{-1}_{KI}$.
Then the resummed cross section is given by
\beq
\sigma^{\rm res}=H_{IJ} \, (R^{-1})^{\dagger}_{JL} (c')^{\dagger}_L
c_K' (R^{-1})_{KI}  \,  e^{E_{KL}} \, ,
\eeq
where the exponent $E_{KL}$ takes contributions at NLL from the
eigenvalues $\lambda_K$ and $\lambda_L$, and is given for $\mu=Q$
by Eq.~(\ref{ELI}).

Note that for the original color bases that we have chosen for all the 
subprocessess, we have $c_I c_J^{\dagger}=0$ for
$I \ne J$; thus, our expressions simplify.
Then the Born cross section can be written as
\beq
\sigma^B=\sum_I H_{II} |c_I|^2 \, ,
\eeq
and the resummed cross section is given by
\beq
\sigma^{\rm res}=\sum_I H_{II} \, (R^{-1})^{\dagger}_{IL} (c')^{\dagger}_L
c_K' (R^{-1})_{KI}  \,  e^{E_{KL}} \, .
\eeq

In the next two subsections we give more detailed results for 
the partonic processes involved in heavy quark production.

\subsection{Diagonalization and numerical results for the 
\protect\newline process $q {\bar q} \rightarrow Q{\bar Q}$}

The Born cross section for a process with a $2\times 2$ anomalous
dimension matrix, $\Gamma_S$, such as $q {\bar q} \rightarrow Q {\bar Q}$,
can be written in a general orthogonal color basis as
\beq
\sigma^B=H_{11}|c_1|^2+H_{22}|c_2|^2\, .
\eeq
The eigenvalues and eigenvectors of $\Gamma_S$
are given by Eqs. (\ref{qqQQev}) and
(\ref{qqQQevec}), respectively.
Then if
$C=(c_1 \quad c_2)$
is the original color basis, the diagonal color basis is
\beq
C' \equiv (c_1' \quad c_2')=CR \, ,
\label{CCpqq}
\eeq
where the columns of $R$ are the eigenvectors of $\Gamma_S$:
\beq
R=\left[e_1 \; e_2\right]=\left[\begin{array}{cc}
\frac{\Gamma_{12}}{\lambda_1-\Gamma_{11}} 
& \frac{\Gamma_{12}}{\lambda_2-\Gamma_{11}} \vspace{2mm} \\
1 & 1 
\end{array}\right] \, .
\eeq
The diagonalized anomalous dimension matrix is then given by
\beq
\Gamma_S^{\rm diag}=R^{-1} \Gamma_S R=\left[\begin{array}{cc}
\lambda_1 & 0 \\
0 & \lambda_2 
\end{array}\right] \, .
\eeq
To write down the Born cross section in the diagonal basis
we use the inverse of Eq.~(\ref{CCpqq}), i.e. $C=C' R^{-1}$,
where the inverse of the matrix $R$ is
\beq
R^{-1}=\frac{(\lambda_1-\Gamma_{11})(\lambda_2-\Gamma_{11})}
{\Gamma_{12}(\lambda_2-\lambda_1)}\left[\begin{array}{cc}
1 & -\frac{\Gamma_{12}}{\lambda_2-\Gamma_{11}} \vspace{2mm} \\
-1 & \frac{\Gamma_{12}}{\lambda_1-\Gamma_{11}}
\end{array}\right] \, .
\eeq
We can thus express the old color basis tensors in terms of the 
new basis.

In addition, we note that the Born cross section for the 
heavy quark production process 
$q{\bar q} \rightarrow Q{\bar Q}$ is pure octet exchange
so that there are further simplifications in the basis
$C=(c_{\rm singlet} \quad c_{\rm octet})$. The Born cross section is then
\beq
\sigma^B_{q{\bar q} \rightarrow Q{\bar Q}}=H_{22}|c_2|^2\, ,
\eeq
with $c_2=c_{\rm octet}$.
Now, $c_{\rm octet}$ can be written in terms of the diagonal basis as 
\beq
c_{\rm octet}=\frac{(\lambda_1-\Gamma_{11}) c_1'
-(\lambda_2-\Gamma_{11})c_2'}{(\lambda_1-\lambda_2)} \, .
\eeq
Then using the above equation we can
rewrite the Born cross section in terms of $|c_1'|^2,|c_2'|^2,c_1'c_2'^*$,
where 
\beq
|c_{1,2}'|^2=\frac{N_c^2 \, \Gamma_{12}^2}{|\lambda_{1,2}-\Gamma_{11}|^2}
+\frac{N_c^2-1}{4}
\eeq
and
\beq
c_1' \, c_2'^*=\frac{N_c^2 \, \Gamma_{12}^2}{(\lambda_1-\Gamma_{11})
(\lambda_2-\Gamma_{11})^*}+\frac{N_c^2-1}{4} \, .
\eeq

To write the resummed cross section in momentum space a prescription
must be chosen to invert the moment space exponentiated cross section.
Here we use the simplest method of Ref.~\cite{LSN}; of course the
diagonalization procedure is general and any prescription for the
inversion can be used.

The resummed partonic cross section is then given by~\cite{NKRV}
\beqa
\sigma^{\rm res}_{q \bar q}(s,m^2)&=&\sum_{i,j=1}^2\int_{-1}^1 d\cos \theta \,
\left[-\int^{s-2ms^{1/2}}_{s_{\rm cut}}
ds_4 f_{q \bar q, ij}(s_4, \theta)
\frac{d{\overline \sigma}_{q \bar q, ij}^{(0)}(s,s_4,\theta)}{ds_4}\right] \, ,
\nonumber \\ 
\label{respart}
\eeqa
where $s_4=s+t_1+u_1$.
The $d{\overline \sigma}_{q \bar q, ij}^{(0)}(s,s_4,\theta)/ds_4$
are components of the differential of the Born cross section which is
given by
\beqa
\frac{d{\overline \sigma}_{q \bar q}^{(0)}(s,s_4,\theta)}{ds_4}&=&
-\pi \alpha_s^2 K_{q {\bar q}} N_c C_F
\frac{1}{4s^4}
\frac{s-s_4}{\sqrt{(s-s_4)^2-4sm^2}}
\nonumber \\ && \hspace{-22mm}
\times \, \left[(3(s-s_4)^2-8sm^2)
(1+\cos^2\theta)+4sm^2(1-\cos^2\theta)\right] \, .
\eeqa
The function $f$ is given at NLL by the exponential
\beq
f_{q \bar q, ij}=\exp[E_{q \bar q, ij}]
=\exp[E_{q \bar q}+E_{q \bar q}(\lambda_i, \lambda_j)] \, ,
\eeq
where in the DIS scheme
\begin{eqnarray}
 E_{q\overline q}^{\rm DIS} &=& \int_{\omega_0}^1\frac{d\omega'}{\omega'}
\Big\{\int_{\omega'^2 Q^2/\Lambda^2}^{\omega' Q^2/\Lambda^2} \frac{d\xi}{\xi}\,
 \Big[ \frac{2 C_F}{\pi} \Big( \alpha_s(\xi)
\nonumber \\ &&  \qquad \qquad
{}+ \frac{1}{2\pi} \alpha^2_s(\xi) K\Big) \Big]
 - \frac{3}{2} \frac{C_F}{\pi} \alpha_s
\Big( \frac{\omega' Q^2}{\Lambda^2}\Big)
\, \Big\} \, ,
\label{Eofm}
\end{eqnarray}
with $\omega_0=s_4/(2m^2)$ and $\Lambda$ the QCD scale parameter.
The color-dependent contribution in the exponent is
\beq
E_{q\overline q}(\lambda_i, \lambda_j)=
-\int_{\omega_0}^1\frac{d\omega'}{\omega'}
\left\{\lambda_i' \left[\alpha_s \left(\frac{\omega'^2 Q^2}
{\Lambda^2}\right), \theta \right]
+{\lambda_j'}^* \left[\alpha_s \left(\frac{\omega'^2 Q^2}{\Lambda^2}\right),
\theta \right] \right\} \, ,
\eeq
in both mass factorization schemes, where $i= 1,2$.
Here $\lambda'=\lambda-\nu^{(f)}/2$ (see Eq.~(\ref{g1g2g3})) where
the $\lambda$'s are the eigenvalues of the soft anomalous dimension matrix.
The cutoff $s_{\rm cut}$ in Eq.~(\ref{respart}) 
regulates the divergence of $\alpha_s$ at low
$s_4$: $s_{\rm cut}>s_{4,{\rm min}}=2m^2\Lambda/Q$.

After some algebra, the resummed partonic cross section at NLL accuracy
is~\cite{NKRV}
\beqa 
&& \hspace{-5mm} 
\sigma^{\rm res}_{q \overline q}(s,m^2) 
= -\sum_{i,j=1}^2\int_{-1}^1 d\cos \theta \, 
\int^{s-2ms^{1/2}}_{s_{\rm cut}} ds_4 \;
\frac{1}{|\lambda_1-\lambda_2|^2} 
\frac{d{\overline \sigma}_{q \overline q}^{(0)}(s,s_4,\theta)}{ds_4}
\nonumber \\ && \hspace {-5mm} \times
\left[\left(\frac{4N_c^2}{N_c^2-1}\Gamma_{12}^2 
+|\lambda_1-\Gamma_{11}|^2 \right) e^{E_{q \overline q, 11}} 
+\left(\frac{4N_c^2}{N_c^2-1}\Gamma_{12}^2 
+|\lambda_2-\Gamma_{11}|^2 \right) e^{E_{q \overline q, 22}}\right.
\nonumber \\ && 
\left. {}-\frac{8N_c^2}{N_c^2-1}\Gamma_{12}^2 
{\rm Re}\left(e^{E_{q \overline q, 12}}\right)
-2{\rm Re} \left((\lambda_1-\Gamma_{11})(\lambda_2-\Gamma_{11})^*
e^{E_{q \overline q, 12}}\right)\right] \, .
\label{partres}
\eeqa
The explicit expressions for the quantities in the above equation are 
long but their derivation is straightforward.
Numerical results for the partonic cross sections (resummed and
one-loop expansions) are given in ~\cite{NKRV}. The one-loop expansion
of the resummed cross section is a good approximation to
the exact NLO result.

The NLL resummed hadronic cross section is given by the convolution
of parton distributions $\phi_{i/h}$, for parton $i$ in hadron $h$,
with the partonic cross section 
\begin{eqnarray}
\sigma^{\rm res}_{q \overline q, {\rm had}}(S,m^2)&=&\sum_{q=u}^b
\int_{\tau_0}^1 d\tau
\int_\tau^1 \frac{dx}{x}
\phi_{q/h_1}(x,\mu^2) \phi_{{\bar q}/h_2}(\frac{\tau}{x},\mu^2)
\sigma_{q \overline q}^{\rm res}(\tau S, m^2) \, , 
\nonumber \\ &&
\label{hadres}
\end{eqnarray}
where $\sigma_{q \overline q}^{\rm res}(\tau S, m^2)$ is defined in 
Eq.~(\ref{partres}) and $\tau_0 = (m + \sqrt{m^2 + s_{\rm cut}})^2/S$.  

Our numerical results for the $t \overline t$ production
cross section at the Fermilab Tevatron with $\sqrt S=1.8$ TeV are shown in
Fig. 5 as functions of the top quark mass.
We use the CTEQ 4D DIS parton densities \cite{CTEQ,PDFLIB}.
Since the parton densities are only available at fixed order, 
the application to a resummed cross section
introduces some uncertainty. 
The NLO exact cross sections, including the factorization scale 
dependence, are shown in Fig.~5 along with the NLO approximate 
cross section, i.e. the one-loop expansion of the resummed cross section,
calculated with $s_{\rm cut}=0$ and $\mu^2=m^2$.  
We note the excellent agreement between the NLO exact and
approximate cross sections.

\begin{figure}
\centerline{
\psfig{file=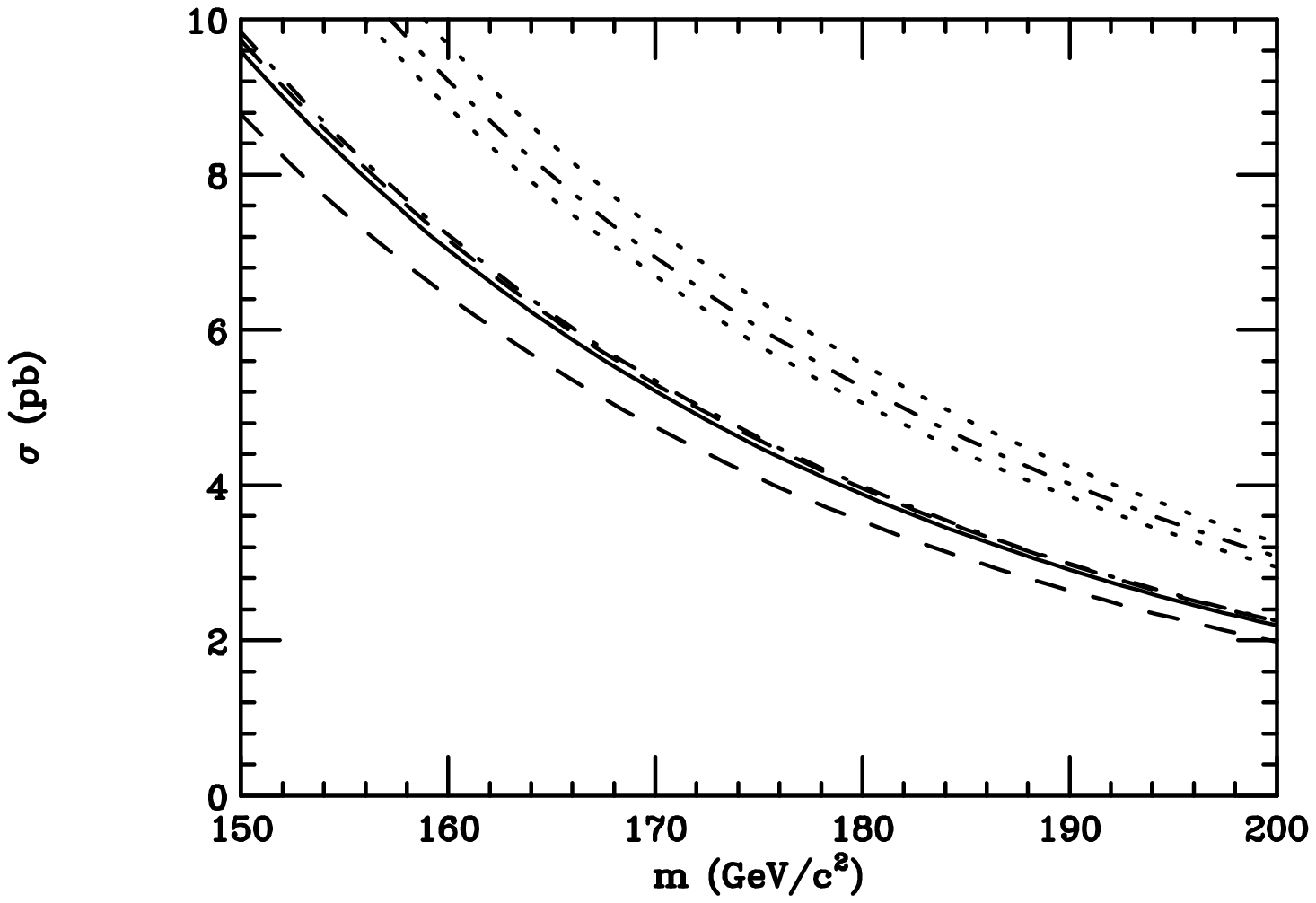,height=3.5in,width=5.05in,clip=}}
{Fig. 5. The NLO exact and approximate and the NLL improved 
hadronic $t \overline t$ production cross sections in 
the $q \overline q$ channel and the DIS scheme are
given as functions of top quark mass for $p \overline p$ collisions at the
Tevatron energy, $\sqrt{S} = 1.8$ TeV. The NLO exact cross section is given
for $\mu^2 = m^2$ (solid curve),
$4m^2$ (lower-dashed) and $m^2/4$ (upper-dashed).
The NLO approximate cross section with $s_{\rm cut} = 0$ is shown for
$\mu^2 = m^2$ (lower dot-dashed).
The NLL improved cross section, Eq.~(\ref{improved}), 
is given for $s_{\rm cut} = 35 s_{4, {\rm  min}}$ (upper dot-dashed), 
$30 s_{4, {\rm min}}$ (upper-dotted) and $40s_{4, {\rm min}}$
(lower-dotted).}
\label{fig5}
\end{figure}

As for the NLL resummed cross section, we note that its scale dependence 
is significantly reduced relative to that of the NLO cross section.
In order to match our results to the exact NLO cross section 
we define the NLL improved cross section
\beq
\sigma_{q \overline q, {\rm had}}^{\rm imp} 
= \sigma_{q \overline q, {\rm had}}^{\rm res} -
\sigma_{q \overline q, {\rm had}}^{\rm NLO, approx} 
+ \sigma_{q \overline q, {\rm had}}^{\rm NLO, exact} \, \, ,
\label{improved} 
\eeq
with the same cut applied to the NLO approximate and the
NLL resummed cross sections.

In Fig. 5 the hadronic improved cross section 
is shown for $\mu^2 = m^2$ 
along with the variation with $s_{\rm cut}$. 
The variation of the improved cross section with change
of cutoff is small over the range 
$30 s_{4, {\rm min}}< s_{\rm cut} <40 s_{4, {\rm min}}$.
At $m = 175$ GeV/$c^2$ and $\sqrt{S} = 1.8$ TeV, 
the value of the improved cross section for 
$q \overline q \rightarrow t \overline t$ with 
$s_{\rm cut}/(2m^2)=0.04$ is 6.0 pb
compared to a NLO cross section of 4.5 pb at $\mu=m$.
At the upgraded Tevatron with $\sqrt{S} = 2$ TeV, the corresponding
value is 7.8 pb compared to a NLO cross section of
5.9 pb at $\mu=m$. We find that 
the corrections relative to NLO are larger for larger scales. 
The $gg$ channel is more complicated and will be discussed in the 
next subsection.  
Adding the $gg$ contribution
we predict a total cross section of 7 pb at $\sqrt{S} = 1.8$ TeV, 
in good agreement with experimental values from CDF, 
$\sigma_{t \overline t}=7.6^{+1.8}_{-1.5}$ pb \cite{CDFn}, and D0, 
$\sigma_{t \overline t}=5.5 \pm 1.8$ pb \cite{D0n}.  
Gluon fusion is more important
for $b$-quark production at HERA-B where threshold resummation
is also of importance \cite{pp,HERAB}.

\subsection{Diagonalization for the process 
$gg \rightarrow {Q \bar Q}$}

The Born cross section for a process with a $3\times 3$ anomalous
dimension matrix, $\Gamma_S$, such as $gg \rightarrow Q{\bar Q}$,
can be written in a general orthogonal color basis as
\beq
\sigma^B=H_{11}|c_1|^2+H_{22}|c_2|^2+H_{33}|c_3|^2\, .
\eeq
The eigenvalues and eigenvectors of $\Gamma_S$ 
are given by Eqs. (\ref{ggQQev}) and
(\ref{ggQQevec}), respectively.
Then if $C=(c_1 \quad c_2 \quad c_3)$
is the original color basis, the diagonal color basis is
\beq
C' \equiv (c_1' \quad c_2' \quad c_3')=CR \, ,
\label{CCpgg}
\eeq
where
\beq
R=\left[e_1 \; e_2 \; e_3 \right]\equiv \left[\begin{array}{ccc}
e_1^1 & e_2^1 & e_3^1 \vspace{2mm} \\
e_1^2 & e_2^2 & e_3^2 \vspace{2mm} \\
1 & 1 & 1
\end{array}\right]
=\left[\begin{array}{ccc}
\frac{\Gamma_{31}}{2(\lambda_1-\Gamma_{11})} &
\frac{\Gamma_{31}}{2(\lambda_2-\Gamma_{11})} &
\frac{\Gamma_{31}}{2(\lambda_3-\Gamma_{11})} \vspace{2mm} \\
\frac{N_c\Gamma_{31}}{4(\lambda_1-\Gamma_{22})} &
\frac{N_c\Gamma_{31}}{4(\lambda_2-\Gamma_{22})} &
\frac{N_c\Gamma_{31}}{4(\lambda_3-\Gamma_{22})} \vspace{2mm} \\
1 & 1 & 1
\end{array}\right] \, ,
\eeq
where, to simplify the expressions below, we denote by $e_i^j$ the
$j$th element of the $i$th eigenvector.
The diagonalized anomalous dimension matrix is
$\Gamma_S^{\rm diag}=R^{-1} \Gamma_S R$.

To write down the Born cross section in the diagonal basis
we use the inverse of Eq.~(\ref{CCpgg}), i.e. $C=C' R^{-1}$,
where the inverse of the matrix $R$ is
\beq
R^{-1}=D^{-1}\left[\begin{array}{ccc}
e_2^2-e_3^2 & e_3^1-e_2^1 & e_2^1 e_3^2-e_3^1 e_2^2 \vspace{2mm} \\
e_3^2-e_1^2 & e_1^1-e_3^1 & e_1^2 e_3^1-e_1^1 e_3^2 \vspace{2mm} \\
e_1^2-e_2^2 & e_2^1-e_1^1 & e_1^1 e_2^2-e_1^2 e_2^1
\end{array}\right]
\eeq
with
\beq
D=(e_1^1-e_3^1)e_2^2-(e_1^1-e_2^1)e_3^2-(e_2^1-e_3^1)e_1^2 \, .
\eeq
Then, the old basis may be expressed in terms of the new basis color tensors,
and we can rewrite the Born cross section in terms of the diagonal basis as
\beq
\sigma^B=\sum_{K,L=1}^3 \sigma^B_{KL} \, ,
\eeq
where $\sigma^B_{KL}$ is the component of the Born cross section
proportional to $c_K' c_L'^*$. The explicit expressions for these components
are, again, long but they are straightforward to derive.
Then the resummed cross section is given by
\beq
\sigma^{\rm res}=\sum_{K,L=1}^3 \sigma^B_{KL} \, e^{E_{KL}} \, ,
\eeq
where the exponent $E_{KL}$ takes contributions at NLL from the 
eigenvalues $\lambda_K$ and $\lambda_L$.

\mysection{Threshold resummation for dijet 
\protect\newline production}

In this section we review the resummation formalism relevant to the
hadronic production of a pair of jets. We will follow the same
methods as for heavy quark production but will encounter
additional complications due to the presence of final state jets.

\subsection{Factorized dijet cross section}
We study dijet production in hadronic processes
\beq
h_a(p_a)+h_b(p_b) \rightarrow J_1(p_1,\delta_1)+J_2(p_2,\delta_2)+X(k)\, ,
\label{dijet}
\eeq
at fixed rapidity interval,
\beq
{\Delta}y={1\over 2}\; \ln\left( {p_1^+\; p_2^- \over p_1^-\; p_2^+}\right)\, ,
\eeq 
with total rapidity, 
\beq
y_{JJ}= {1\over 2}\; \ln\left( {p_1^++p_2^+ \over p_1^-+p_2^-}\right)\, .
\eeq
The jets are defined by cone angles $\delta_1$ and $\delta_2$.
The introduction of cones
removes all the final-state  collinear singularities 
from the partonic cross section, which
is then infrared safe, once the initial-state collinear singularities 
have been factored into universal parton
distribution functions.
We shall assume that the cones are small enough so that 
contributions proportional to $\delta_i\ll 1$ may be neglected,
but large enough so that
$\alpha_s(Q)\ln(1/\delta_i) \ll 1$,
where $Q$ is any of the hard scales of the 
cross section, such as  the momentum transfer \cite{GS78,gsbook}.

To construct the dijet cross sections, we define a large invariant,
$M_{JJ}$, which is held fixed. A natural choice is
the dijet invariant mass,
\beq
M^2_{JJ}=(p_1+p_2)^2\, ,
\label{jetmass}
\eeq
which is the analog of $Q^2$ for heavy
quark production, but other choices are possible, for example
the scalar product of the  two jet momenta
\beq
M^2_{JJ}=2p_1 \cdot p_2 \, .
\label{jetprod}
\eeq
In both cases, large
$M_{JJ}$ at fixed $\Delta y$ implies a large momentum transfer
in the partonic subprocess.
The nature of the resummed cross section depends critically 
on this choice. As we shall see, the leading behavior of the resummed 
cross section is the same as for Drell-Yan or heavy quark production
for the first choice for $M_{JJ}$, Eq.~(\ref{jetmass}), while it is
different for the second choice, Eq.~(\ref{jetprod}). 

The dijet cross section is given in factorized form by   
\beqa
\frac{d\sigma_{h_ah_b{\rightarrow}J_1J_2}(S,M_{JJ},y,\Delta y,
\delta_1,\delta_2)}{dM^2_{JJ}\; dy_{JJ} \; d{\Delta}y}&=& 
\sum_{f_a,f_b=q,\overline{q},g} 
\int \frac{dx_a}{x_a} \, \frac{dx_b}{x_b}\; \phi_{f_a/h_a}(x_a,\mu^2)
\nonumber \\ &&  \hspace{-40mm}
\times \, \phi_{f_b/h_b}(x_b,\mu^2) \; 
H_{f_af_b}\left(\frac{M_{JJ}^2}{x_ax_bS},y,\Delta y,\frac{M_{JJ}}{\mu}, 
\alpha_s(\mu^2),\delta_1,\delta_2\right) ,
\nonumber \\
\eeqa
where again $H$ is the hard scattering and the $\phi$'s are parton
distribution functions.
The threshold for the partonic subprocess is given 
in terms of the variable $z$,
\beq
z=\frac{M^2_{JJ}}{x_ax_b{S}}=\frac{M^2_{JJ}}{s}\, ,
\eeq
where, as before, $S=(p_a+p_b)^2$ and $s=x_ax_bS$.  
At $z_{\rm max}=1$ (partonic threshold) there is  just enough 
partonic energy to produce the observed final state, 
with no additional radiation.
The lower limit of $z$ is
\beq
z_{\rm min}\equiv\tau=\frac{M^2_{JJ}}{S} \, .
\eeq

We may rewrite the cross section, introducing an explicit 
integration over $z$, as
\beqa
&&\frac{d\sigma_{h_ah_b{\rightarrow}J_1J_2}(S,M_{JJ},y,\Delta y,
\delta_1,\delta_2)}
{dM^2_{JJ} \; dy_{JJ} \; d{\Delta}y}= 
\sum_{f_a,f_b=q,{\bar q},g}\int_{\tau}^1dz
\int \frac{dx_a}{x_a} \, \frac{dx_b}{x_b}
\nonumber \\
&&\hspace{15mm}\times \, 
\phi_{f_a/h_a}(x_a,\mu^2) \;\phi_{f_b/h_b}(x_b,\mu^2)\; 
\delta\left(z-\frac{M^2_{JJ}}{s}\right) \;
\delta\left(y_{JJ}-\frac{1}{2}\ln\frac{x_a}{x_b}\right)
\nonumber \\
&&\hspace{15mm}\times
\sum_{f_1,f_2=q, {\bar q},g}
\hat{\sigma}_{f_af_b\rightarrow f_1f_2}\left(1-z,\frac{M_{JJ}}{\mu},
{\Delta}y,\alpha_s(\mu^2),\delta_1,\delta_2\right)\, ,
\eeqa
where again we have used the observation~\cite{LaSt} that we can treat 
the total rapidity of the dijets  as a constant, equal to its value
at threshold. Thus we now have a simplified hard scattering function, 
$\hat\sigma_{f_af_b\rightarrow f_1f_2}$.

To calculate $\hat\sigma_{f_af_b\rightarrow f_1f_2}$ 
at any order in perturbation theory, we construct the partonic cross section
for the process $f_a+f_b \rightarrow f_1+f_2$, as above,
\beqa
&& \frac{d\sigma_{f_af_b{\rightarrow}J_1J_2}
(S,M_{JJ},\Delta y,\delta_1,\delta_2)}
{dM^2_{JJ} \; d{\Delta}y}=\int_{\tau}^1 dz
\int \frac{dx_a}{x_a} \, \frac{dx_b}{x_b}\; \phi_{f_a/f_a}(x_a,\mu^2) 
\nonumber \\ && \times \,
\phi_{f_a/f_a}(x_a,\mu^2) \;
\phi_{f_b/f_b}(x_b,\mu^2) \, 
\delta\left(z-\frac{M^2_{JJ}}{s}\right) 
\nonumber \\ && \times \,
\sum_{f_1,f_2=q,{\bar q},g}
\hat{\sigma}_{f_af_b\rightarrow f_1f_2}
\left(1-z,\frac{M_{JJ}}{\mu},{\Delta}y,\alpha_s(\mu^2),
\delta_1,\delta_2\right)\, ,
\label{sigpart}
\eeqa
where we have integrated over the total rapidity. We then
factorize the initial-state collinear divergences into the 
light-cone  distribution functions $\phi_{f/f}$, expanded to the
same order in $\alpha_s$, and thus obtain
the perturbative expansion for the infrared-safe hard
scattering function, $\hat {\sigma}$. 
As for heavy quark production, again we note that the 
leading power as $z \rightarrow 1$ comes entirely from flavor diagonal 
parton distributions. Moreover, we may sum, at leading power, over the
flavors of the final-state partons $f_1,f_2$. 

By taking a Mellin transform of the rapidity-integrated partonic cross section 
(\ref{sigpart}) with respect to $\tau$, 
the above convolution becomes a product
\beqa
\int_0^1 d\tau\; \tau^{N-1}\; 
\frac{d\sigma_{f_af_b{\rightarrow}J_1J_2}
(S,M_{JJ},\Delta y,\delta_1,\delta_2)}
{dM^2_{JJ} \; d{\Delta}y} 
&=& \sum_{\a}{\tilde \phi}_{f_a/f_a}(N,\mu^2,\epsilon)\; 
\nonumber \\ && \hspace{-60mm} \times \,
{\tilde \phi}_{f_b/f_b}(N,\mu^2,\epsilon) \;
{{\tilde{\sigma}}}_{\a}(N,M_{JJ}/\mu,\alpha_s(\mu^2),\delta_1,
\delta_2)\, ,
\label{moment}
\eeqa
where ${\a}$ denotes the partonic processes $f_a+f_b{\rightarrow}f_1+f_2$,
and with $\tilde{\sigma}_{\a}(N)=\int_0^1dz\; z^{N-1}\hat{\sigma}_{\a}(z)$, 
and $\tilde{\phi}(N)=\int_0^1dx\; x^{N-1}\phi(x)$, as before.  

\begin{figure}
\centerline{
\psfig{file=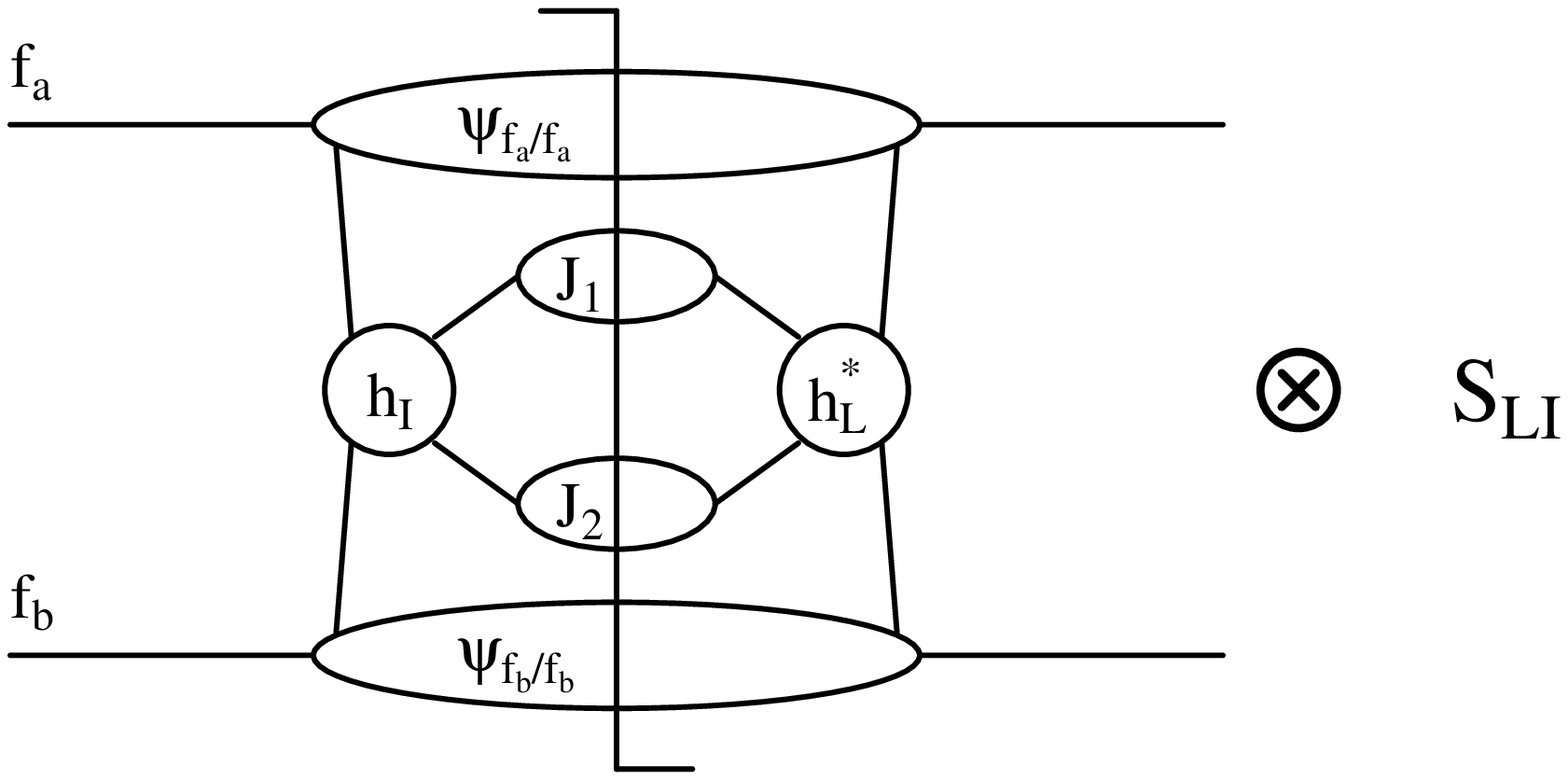,height=2.5in,width=5.05in,clip=}}
{Fig. 6. Refactorization for dijet production. The soft gluon function 
is as in Fig. 1(b).}
\label{fig6}
\end{figure}

We can refactorize the cross section, 
as shown in Fig.~6, into hard components $H_{IL}$,
which describe the truly short-distance hard-scattering,
center-of-mass distributions $\psi$, associated with gluons
collinear to the incoming partons, a soft gluon function
$S_{LI}$ associated with soft gluons, and jet functions $J_i$,
associated with gluons collinear to the outgoing jets.
As before, $I$ and $L$ are color indices that describe the color structure
of the hard scattering. 
The refactorized cross section may then be written as
\beqa
&&\hspace{-10mm}\int_0^1 d\tau\; \tau^{N-1}\;
\frac{d\sigma_{f_af_b{\rightarrow}J_1J_2}
(S,M_{JJ},\Delta y,\delta_1,\delta_2)}
{dM^2_{JJ} \; d{\Delta}y} 
\nonumber\\ && 
=\sum_{\a}\sum_{IL} 
H_{IL}^{(\a)}\left({M_{JJ}\over\mu},\Delta y,\alpha_s(\mu^2)\right) \;
{\tilde S}_{LI}^{(\a)} \biggl ( {M_{JJ}\over N \mu},
\Delta y,\alpha_s(\mu^2) \biggr)
\nonumber\\ && \quad\times\; {\tilde\psi}_{f_a/f_a}\left(N,{M_{JJ}\over \mu},
\alpha_s(\mu^2),\epsilon \right) \;
{\tilde\psi}_{f_b/f_b}\left(N,{M_{JJ}\over \mu },
\alpha_s(\mu^2),\epsilon \right)
\nonumber \\ && \quad \times \; 
{\tilde J}_{(f_1)}\left(N,{M_{JJ}\over \mu },
\alpha_s(\mu^2), \delta_1\right)\; 
{\tilde J}_{(f_2)}\left(N,{M_{JJ}\over \mu },\alpha_s(\mu^2),\delta_2\right)
\nonumber \\ && \quad \quad
{}+{\cal O}(1/N)\, .  
\label{refact}
\eeqa

This refactorization is similar to the heavy quark 
case except that now we have to include in addition  
outgoing jet functions in order to absorb
the final state collinear singularities (the mass of the heavy
quarks eliminates final state collinear singularities in heavy quark 
production). Expressions for the hard components $H_{IL}$
and the center-of-mass distributions $\psi$ were given in Section 2
in the context of heavy quark production.

The soft function again represents the coupling of soft gluons
to the partons in the hard scattering; this coupling, 
as we saw in Section 2.2, is described by eikonal lines.
The construction of the eikonal cross section for dijet production
is similar to the heavy quark case, apart from the outgoing jets.
The collinear dynamics of the eikonal final-state
jets are summarized by the matrix elements~\cite{KOS1} 
\beqa       
j^{(f_i)}_{{\rm{OUT}}}\left({w_iM_{JJ}\over \mu},\alpha_s(\mu^2),
\delta_i\right) &=&
\sum_{\xi}\; \delta\left(w_i-w(\xi,\delta_i)\right)
\nonumber\\ && \hspace{-40mm} \times \, 
\langle 0|\; {\rm Tr}\bigg\{\; {\bar T}
[\Phi^{(f_i)}_{\beta_i}{}^\dagger(\infty,0;0)]|\xi \rangle \langle \xi|
T[\Phi^{(f_i)}_{\beta_i}(\infty,0;0)]\; \bigg\}\; |0\rangle\, , 
\label{eikoutjet}
\eeqa
where the ordered exponentials $\Phi_{\beta}$ are defined in
Eq.~(\ref{ordexp}), and with $i=1,2$ 
and $\xi$ a set of intermediate states which contribute
to the weight $w_i$. 

We then construct moments of  the soft function by dividing the moments of 
the eikonal cross section (defined in analogy 
to Eq.~(\ref{sigeik}), see Ref.~\cite{KOS1}) 
by the product of moments of the eikonal jets, 
Eqs.\ (\ref{eikinjet}) and (\ref{eikoutjet}),
\beqa
{\tilde S}^{(\a)}_{LI}\left({M_{JJ}\over N\mu},\Delta y,\alpha_s(\mu^2)\right)
&=&
{{\tilde \sigma_{LI}}^{(\a,{\rm eik})}\left({M_{JJ}\over N\mu},
\Delta y,\alpha_s(\mu^2),\epsilon\right)
\over
{\tilde j}^{(f_a)}_{\rm{IN}}\left({M_{JJ}\over N\mu},\alpha_s(\mu^2),
\epsilon\right)\; 
{\tilde j}^{(f_b)}_{\rm{IN}}\left({M_{JJ}\over N\mu},\alpha_s(\mu^2),
\epsilon\right)}
\nonumber\\ && \hspace{-10mm} \times \,
\frac{1}
{{\tilde{j}}^{(f_1)}_{{\rm{OUT}}}\left(\frac{M_{JJ}}{N\mu},
\alpha_s(\mu^2),\delta_1\right) \: 
{\tilde{j}}^{(f_2)}_{{\rm{OUT}}}\left(\frac{M_{JJ}}{N\mu},
\alpha_s(\mu^2),\delta_2\right)}\, .
\nonumber\\
\label{sigeikDY}
\eeqa
                         
\subsection{Resummed dijet cross section}

Comparing Eqs.\ (\ref{moment}) and  (\ref{refact}), 
the refactorized expression for the Mellin transform of 
the hard scattering function is
\beqa
\tilde{{\sigma}}_{\a}(N)&=&
\left[\frac{ {\tilde{\psi}}_{f_a/f_a}(N,M_{JJ}/\mu,\epsilon) \,
{\tilde{\psi}}_{f_b/f_b}(N,M_{JJ}/\mu,\epsilon)}
{{\tilde{\phi}}_{f_a/f_a}(N,\mu^2,\epsilon) \, 
{\tilde{\phi}}_{f_b/f_b}(N,\mu^2,\epsilon)}
 \right]\nonumber \\ && \hspace{-15mm} \times \,
\sum_{IL}H_{IL}^{(\a)}\left({M_{JJ}\over\mu},\Delta y,\alpha_s(\mu^2)\right)\; 
{\tilde S}_{LI}^{(\a)} \left( {M_{JJ}\over N \mu},\Delta y,\alpha_s(\mu^2) 
\right)
\nonumber \\ && \hspace{-15mm} \times \,
{\tilde J}_{(f_1)}\left(N,{M_{JJ}\over \mu},\alpha_s(\mu^2),\delta_1\right)\; 
{\tilde J}_{(f_2)}\left(N,{M_{JJ}\over \mu},\alpha_s(\mu^2),\delta_2\right)\, .
\label{sigfactjet}
\eeqa

We discussed resummation for the ratio $\psi/\phi$ 
and the soft gluon function in the context of heavy quark 
production in Section 2. 
Thus, all we need to write down the resummed dijet cross section are
expressions for the resummation of the final state jets. 

The moments of the final-state jet with
$M_{JJ}^2=2p_1\cdot p_2$ are given by~\cite{KOS1}
\beq
\tilde{J}_{(f_i)}\left(N,{M_{JJ}\over \mu},\alpha_s(\mu^2),\delta_i\right)
=\exp \left[E'_{(f_i)}(N,M_{JJ})\right]\, ,
\label{finaljet}
\eeq
with
\beqa
E'_{(f)}\left(N,M_{JJ}\right)
&=&
\int^1_0 dz \frac{z^{N-1}-1}{1-z}\; 
\left \{\int^{(1-z)}_{(1-z)^2} \frac{d\lambda}{\lambda} 
A^{(f)}\left[\alpha_s(\lambda M_{JJ}^2)\right] \right.
\nonumber\\ &&  \quad \quad \left.
{}+B'_{(f)}\left[\alpha_s((1-z) M_{JJ}^2) \right] \right\}\, , 
\label{Eprexp}
\eeqa
where the function $A^{(f)}$ is the same as in Eq.\ (\ref{Aexp})
and the lowest-order term in $B'_{(f)}$
may be read off from the one-loop jet function.  
The results include a gauge dependence, which cancels
against a corresponding dependence in the soft anomalous
dimension matrix.
The leading logarithms for final-state jets with $M_{JJ}^2=2p_1\cdot p_2$
are negative and give a suppression to the cross section, in contrast
to the initial-state leading-log contributions.

The moments of the final-state jet with
$M_{JJ}^2=(p_1+p_2)^2$ are given by Eq.~(\ref{finaljet}) with~\cite{KOS1}
\beq
E'_{(f_i)}(N,M_{JJ})=
\int_\mu^{M_{JJ}/N} {d\mu' \over \mu'}\; \; 
C'_{(f_i)}(\alpha_s(\mu'{}^2))\, ,
\label{Epr2exp}
\eeq
where the first term in the series for $C'_{(f_i)}(\alpha_s)$ 
may be read off from a one-loop calculation. 
The leading logarithmic behavior of the 
cross section in this case is not affected by the final state jets, 
so we always have an enhancement of the cross section at leading logarithm, 
as is the case for Drell-Yan and heavy quark cross sections.  

Using Eqs.~(\ref{sigfactjet}), (\ref{psiphimu}), 
(\ref{rgesol}), and (\ref{finaljet}),  
we can write the resummed dijet cross section in moment space as
\beqa
\tilde{{\sigma}}_{\a}(N) &=& R_{(f)}^2\; 
\exp \left \{ \sum_{i=a,b} \left[ E^{(f_i)}(N,M_{JJ}) \right. \right. 
\nonumber\\
&\ & \hspace{20mm} \left. \left.
{}-2\int_\mu^{M_{JJ}}{d\mu'\over\mu'}\; 
\left [\gamma_{f_i}(\alpha_s(\mu'{}^2))-\gamma_{f_if_i}(N,\alpha_s(\mu'{}^2)) 
\right] \right] \right\}
\nonumber \\
&\ &\times\; \exp \left \{\sum_{j=1,2}E'_{(f_j)}(N,M_{JJ}) \right\}
\nonumber\\ && \times\; {\rm Tr} \left\{ 
H^{(\a)}\left({M_{JJ}\over\mu},\Delta y,\alpha_s(\mu^2)\right) \right.
\nonumber\\ &&\times\;
\bar{P} \exp \left[\int_\mu^{M_{JJ}/N} {d\mu' \over \mu'}\; 
\Gamma_S^{(\a)}{}^\dagger\left(\alpha_s(\mu'^2)\right)\right]\;
{\tilde S}^{(\a)} \left(1,\Delta y,\alpha_s\left(M_{JJ}^2/N^2\right) \right)
\nonumber\\ && \left. \times\; 
P \exp \left[\int_\mu^{M_{JJ}/N} {d\mu' \over \mu'}\; \Gamma_S^{(\a)}
\left(\alpha_s(\mu'^2)\right)\right] \right\}\, ,
\eeqa
where $E^{(f_i)}$ is given by Eq.~(\ref{Eexp}),
and $E'_{(f_j)}$ is given by  Eq.~(\ref{Eprexp}) or Eq.~(\ref{Epr2exp}).
This expression is similar to Eq.~(\ref{resHQ}) 
for heavy quark production except for the
addition of the exponents for the final-state jets. 

We give explicit expressions for the soft anomalous
dimensions for the partonic processes relevant to dijet
production in the next three sections.

\mysection{Soft anomalous dimension matrices for 
\protect\newline processes involving quarks}

In this section we present results for the soft anomalous dimension
matrices for the processes $q {\bar q} \rightarrow q {\bar q}$,
$qq \rightarrow qq$, and ${\bar q} {\bar q} \rightarrow {\bar q} {\bar q}$.
The matrices for all these processes are $2 \times 2$. Hence the
general diagonalization procedure for these processes
follows along the same lines as we discussed at the beginning 
of Section 5.2.

Since the results are somewhat lengthy, we first introduce  some notation
which will simplify the expressions for $\Gamma_S$.

\subsection{Notation}
We consider partonic processes 
$f_a\left(p_a, r_a \right)+f_b\left(p_b, r_b \right)$ $\rightarrow$ 
$f_1\left(p_1, r_1 \right)+f_2\left(p_2, r_2 \right)$,
where the $r_i$ are color labels, and the $p_i$ are momenta. 
To facilitate the presentation of the results for $\Gamma_S$ we
introduce the notation
\beq
\T\equiv \ln\left(\frac{-t}{s}\right)+\pi i \, , \quad \quad
\U\equiv \ln\left(\frac{-u}{s}\right)+\pi i \, ,
\label{eq:new2form}
\eeq
where
\beq
s=\left(p_a+p_b \right)^2 \, , \quad t=\left(p_a-p_1 \right)^2 \, , 
\quad u=\left(p_a-p_2 \right)^2 \, ,
\label{Mandlst}
\eeq
are the usual Mandelstam invariants.
Concerning the choice of physical channel $s$, $t$, or $u$, for the
definition of the color basis, we note that we can choose any channel. 
For the processes $q {\bar q} \rightarrow q {\bar q}$,
$qq \rightarrow qq$ and ${\bar q} {\bar q} \rightarrow {\bar q} {\bar q}$,
$qg \rightarrow qg$ and ${\bar q}g\rightarrow {\bar q}g$,
as well as the process $gg \rightarrow gg$
we will use $t$-channel bases, which seem to be the
natural choice when analyzing forward scattering \cite{SoSt}.
The processes $q {\bar q}\rightarrow gg$ and $gg \rightarrow q{\bar q}$  
are better described in terms of $s$-channel 
color structures, so we will give results for them
in $s$-channel color bases.

Since the full cross section is gauge independent, i.e. independent of the
choice of the axial gauge-fixing vector $n^{\mu}$,
the gauge dependence in the product of the hard and soft functions,
$H^{(\a)} \, S^{(\a)}$, 
must cancel the gauge dependence of the incoming and outgoing jets, $\psi$ 
and $J_{(f_i)}$.
Now, the jets are incoherent relative to the hard and soft functions, 
so the gauge dependence
of the anomalous dimension matrices $\Gamma_S^{(\a)}$ must
be proportional to the identity matrix.
Then we can rewrite the anomalous dimension matrix as
\beq
(\Gamma^{(\a)}_S)_{KL}=(\Gamma^{(\a)}_{S'})_{KL}
+ \delta_{KL}  \frac{\alpha_s}{\pi} \sum_{i=a,b,1,2}C_{(f_i)} \, 
\frac{1}{2} \,  (-\ln\nu_i-\ln 2+1-\pi i) \, ,
\label{gammagaug}
\eeq
with $C_{f_i}=C_F\ (C_A)$ for a quark (gluon), and with
$\nu_i \equiv (v_i \cdot n)^2/|n|^2$, as in Eq. (\ref{nui}).
Here the dimensionless and lightlike velocity vectors $v_i^{\mu}$ are
defined by
\beq
p_i^{\mu}=\frac{M_{JJ}}{\sqrt{2}} v_i^{\mu} \, , \quad \quad i=a,b,1,2 \, ,
\eeq
and satisfy $v_i^2=0$.

For the subprocesses involved in dijet production we will present
the explicit expressions for $\Gamma^{(\a)}_{S'}$. The full anomalous
dimension matrix can then be retrieved from Eq.~(\ref{gammagaug}).

We now give the results for the anomalous 
dimension matrices 
$\Gamma^{(\a)}_{S'}$ for partonic processes involving quarks.

\subsection{Soft anomalous dimension for $q \bar{q}\rightarrow q \bar{q}$}

First, we present the $2 \times 2$  soft 
anomalous dimension matrix for the process
\beq
q\left(p_a, r_a \right)+\bar{q}\left(p_b, r_b \right) \rightarrow
q\left(p_1, r_1 \right)+\bar{q}\left(p_2, r_2 \right) \, ,
\eeq
in the $t$-channel singlet-octet color basis
\beqa
c_1&=&\delta_{r_a r_1}\delta_{r_b r_2} \, , \nonumber\\
c_2&=&(T_F^c)_{r_1 r_a}(T_F^c)_{r_b r_2} 
=-\frac{1}{2N_c}\delta_{r_a r_1}\delta_{r_b r_2}+\frac{1}{2}
\delta_{r_a r_b} \delta_{r_1 r_2}.
\eeqa
We find the soft anomalous dimension matrix~\cite{BottsSt,KOS2}
\beq
\Gamma_{S'}=\frac{\alpha_s}{\pi}\left[
                \begin{array}{cc}
                 2{C_F}\T  &   -\frac{C_F}{N_c} \U  \vspace{2mm} \\
                -2\U    &-\frac{1}{N_c}(\T-2\U)
                \end{array} \right]\, .
\eeq
The dependence on the logarithmic ratio $\T$ is diagonal in 
this $t$-channel color basis.
In the forward region of the partonic scattering 
($\T\rightarrow -\infty$), where $\Gamma_{S}$ becomes diagonal,
color singlet exchange is exponentially enhanced relatively to color octet 
\cite{SoSt}.
The general forms of the eigenvalues and eigenvectors 
of $\Gamma_S$ are given by
Eqs.~(\ref{qqQQev}) and (\ref{qqQQevec}); 
explicit expressions are given in Ref.~\cite{KOS2}.

Finally we note that this result for $\Gamma_S$ 
is consistent with the massless limit 
of the anomalous dimension matrix for the heavy quark production
process $q{\bar q} \rightarrow Q{\bar Q}$ \cite{Thesis,KS}.

\subsection{Soft anomalous dimension for $q q\rightarrow q q$
and ${\bar q}{\bar q}\rightarrow {\bar q}{\bar q}$}
Next, we consider the process
\beq 
q\left(p_a, r_a \right)+q\left(p_b, r_b \right) \rightarrow
q\left(p_1, r_1 \right)+q\left(p_2, r_2 \right) \, ,
\eeq
in the $t$-channel singlet-octet color basis
\beqa
c_1&=&(T_F^c)_{r_1 r_a}(T_F^c)_{r_2 r_b} 
=-\frac{1}{2N_c}\delta_{r_a r_1} \delta_{r_b r_2}
+\frac{1}{2} \delta_{r_a r_2} \delta_{r_b r_1} \, ,
\nonumber\\
c_2&=&\delta_{r_a r_1} \delta_{r_b r_2}.
\eeqa
The anomalous dimension matrix is given by~\cite{BottsSt,KOS2}
\beq
\Gamma_{S'}=\frac{\alpha_s}{\pi}\left[
                \begin{array}{cc}
                -\frac{1}{N_c}(\T+\U)+2C_F \U  &  2\U \vspace{2mm} \\
                 \frac{C_F}{N_c} \U    & 2{C_F}\T
                \end{array} \right].
\eeq
Again, the dependence on the logarithmic ratio $\T$ is diagonal in this
$t$-channel color basis, and the color singlet dominates the octet 
in the forward region of the partonic scattering ($\T\rightarrow -\infty$),
where $\Gamma_{S}$ becomes diagonal.

Note that the same anomalous dimension matrix applies to the process
\beq 
{\bar q}\left(p_1, r_1 \right)+{\bar q}\left(p_2, r_2 \right) \rightarrow
{\bar q}\left(p_a, r_a \right)+{\bar q}\left(p_b, r_b \right) \, .
\eeq

Again, the general forms of the eigenvalues and eigenvectors 
of $\Gamma_S$ are given by Eqs.~(\ref{qqQQev}) and (\ref{qqQQevec}). 
Explicit expressions for the eigenvalues and eigenvectors are given in 
Ref.~\cite{KOS2}.

\mysection{Soft anomalous dimension matrices for 
\protect\newline processes involving quarks and gluons}

Here we present the soft anomalous dimension matrices for the processes
$q \bar{q}\rightarrow g g$, $g g \rightarrow q \bar{q}$, 
$qg \rightarrow qg$, and $\bar{q} g \rightarrow \bar{q} g$.
The matrices for all these processes are $3 \times 3$.
Moreover, we shall see that they are of the same general form 
as the anomalous dimension matrix
for the heavy quark production process $gg \rightarrow Q {\bar Q}$.
Hence the general diagonalization procedure for these processes
follows the same lines as we discussed in Section 5.3
in the context of heavy quark production in the channel 
$gg \rightarrow Q {\bar Q}$.

\subsection{Soft anomalous dimension for $q \bar{q}\rightarrow g g$ and
$g g \rightarrow q \bar{q}$}

First, we present the soft anomalous dimension matrix for 
the process
\beq
q\left(p_a, r_a \right)+\bar{q}\left(p_b, r_b \right) \rightarrow
g\left(p_1, r_1 \right)+g\left(p_2, r_2 \right) \, ,
\eeq
in the $s$-channel color basis
\beq
c_1=\delta_{r_a r_b}\delta_{r_1 r_2} \, , \quad
c_2=d^{r_1 r_2 c}{\left( T_F^c \right)}_{r_b r_a} \, , \quad
c_3=if^{r_1 r_2 c}{\left( T_F^c \right)}_{r_b r_a} \, .
\eeq
We find~\cite{KOS2} 
\beq
\Gamma_{S'}=\frac{\alpha_s}{\pi}\left[
                \begin{array}{ccc}
                 0  &   0  & \U-\T  \vspace{2mm} \\ 
                 0  &   \frac{C_A}{2}\left(\T+\U \right)    & \frac{C_A}{2}
\left(\U-\T\right) \vspace{2mm} \\ 
                 2\left(\U-\T \right)  & \frac{N_c^2-4}{2N_c}\left(\U-\T 
\right)  & \frac{C_A}{2}\left(\T+\U \right)
                \end{array} \right].
\label{Gammaqqgg}
\eeq
The same anomalous dimension describes also the time-reversed process
\cite{Thesis,KS}
\beq
g\left(p_1, r_1 \right)+g\left(p_2, r_2 \right) \rightarrow
\bar{q}\left(p_a, r_a \right)+q\left(p_b, r_b \right) .
\eeq
We note that the anomalous dimension matrix, Eq.~(\ref{Gammaqqgg}),
is of the general form of Eq.~(\ref{GammaggQQ33}).
Then the general forms of its eigenvalues and eigenvectors are given by
Eqs.~(\ref{ggQQev}) and (\ref{ggQQevec}); 
explicit expressions are given in Ref.~\cite{KOS2}.

Finally we note that this result for $\Gamma_{S}$ is consistent 
with the massless limit of the anomalous dimension matrix for 
the heavy quark production process $gg \rightarrow Q{\bar Q}$
\cite{Thesis,KS}.

\subsection{Soft anomalous dimension for $qg \rightarrow qg$ and
$\bar{q} g \rightarrow \bar{q} g$}

Next, we consider the ``Compton'' process
\beq
q\left(p_a, r_a \right)+g\left(p_b, r_b \right) \rightarrow
q\left(p_1, r_1 \right)+g\left(p_2, r_2 \right) \, .
\eeq
The soft anomalous dimension matrix 
in the $t$-channel color basis
\beq
c_1=\delta_{r_a r_1}\delta_{r_b r_2} \, , \quad
c_2=d^{r_b r_2 c}{\left( T_F^c \right)}_{r_1 r_a} \, , \quad
c_3=if^{r_b r_2 c}{\left( T_F^c \right)}_{r_1 r_a} \, ,
\label{eq:basqgqg}
\eeq
is given by \cite{KOS2}
\beq
\Gamma_{S'}=\frac{\alpha_s}{\pi}\left[
                \begin{array}{ccc}
                 \left( C_F+C_A \right) \T  &   0  & \U  \vspace{2mm} \\ 
                 0  &   C_F \T+ \frac{C_A}{2} \U     & \frac{C_A}{2} \U  
\vspace{2mm} \\
                 2\U  & \frac{N_c^2-4}{2N_c}\U  &  C_F \T+ \frac{C_A}{2}\U
                \end{array} \right] \, ,
\label{Gammaqgqg}
\eeq
which also applies to the process
\beq
\bar{q}\left(p_1, r_1 \right)+g\left(p_2, r_2 \right) \rightarrow
\bar{q}\left( p_a, r_a \right)+g\left(p_b, r_b \right) \, .
\eeq
The dependence on
the logarithmic ratio $\T$ is diagonal in
this $t$-channel color basis, and the $t$-channel
color singlet dominates in the forward region ($\T \rightarrow -\infty$)
where $\Gamma_{S}$ becomes diagonal.
Again, we note that the anomalous dimension matrix, Eq.~(\ref{Gammaqgqg}),
is of the general form of Eq.~(\ref{GammaggQQ33}), so
the general forms of its eigenvalues and eigenvectors are given by
Eqs.~(\ref{ggQQev}) and (\ref{ggQQevec}).
Explicit expressions for the eigenvalues and eigenvectors
are given in Ref.~\cite{KOS2}.

\mysection{Soft anomalous dimension matrix and 
\protect\newline diagonalization for $gg \rightarrow gg$}

\subsection{Soft anomalous dimension matrix for $gg \rightarrow gg$}

The final, and by far more complicated, process that we consider is
\beq
g\left(p_a, r_a \right)+g\left(p_b, r_b \right) \rightarrow
g\left(p_1, r_1 \right)+g\left(p_2, r_2 \right) \, .
\eeq
The choice of a color basis, in which a four-gluon diagram can be expanded
\cite{Macfar,Dixon}, is more difficult in this case.
In Ref.~\cite{KOS2} an initial overcomplete color basis of nine elements 
was found convenient for the calculations, resulting in a $9 \times 9$
anomalous dimension matrix. This was then reduced in a complete color basis
to an $8 \times 8$ matrix.

A complete color basis for the process $gg \rightarrow gg$
is given by the eight color structures~\cite{KOS2}
\beqa
c_1&=&\frac{i}{4}\left[f^{r_a r_b l}
d^{r_1 r_2 l} - d^{r_a r_b l}f^{r_1 r_2 l}\right] \, ,
\nonumber \\ 
c_2&=&\frac{i}{4}\left[f^{r_a r_b l}
d^{r_1 r_2 l} + d^{r_a r_b l}f^{r_1 r_2 l}\right] \, ,
\nonumber \\ 
c_3&=&\frac{i}{4}\left[f^{r_a r_1 l}
d^{r_b r_2 l}+d^{r_a r_1 l}f^{r_b r_2 l}\right] \, , 
\nonumber \\
c_4&=&P_1(r_a,r_b;r_1,r_2)=\frac{1}{8}\delta_{r_a r_1} 
\delta_{r_b r_2} \, ,
\nonumber \\
c_5&=&P_{8_S}(r_a,r_b;r_1,r_2)
=\frac{3}{5} d^{r_ar_1c} d^{r_br_2c} \, ,
\nonumber \\
c_6&=&P_{8_A}(r_a,r_b;r_1,r_2)=\frac{1}{3} f^{r_ar_1c} f^{r_br_2c} \, ,
\nonumber \\
c_7&=&P_{10+{\overline {10}}}(r_a,r_b;r_1,r_2)=    
\frac{1}{2}(\delta_{r_a r_b} \delta_{r_1 r_2}
-\delta_{r_a r_2} \delta_{r_b r_1})
-\frac{1}{3} f^{r_ar_1c} f^{r_br_2c} \, ,
\nonumber \\
c_8&=&P_{27}(r_a,r_b;r_1,r_2)=\frac{1}{2}(\delta_{r_a r_b} 
\delta_{r_1 r_2} +\delta_{r_a r_2} \delta_{r_b r_1})
-\frac{1}{8}\delta_{r_a r_1} \delta_{r_b r_2}
\nonumber \\ &&
-\frac{3}{5} d^{r_ar_1c} d^{r_br_2c} \, ,
\label{8x8basis}
\eeqa
where the $P$'s are $t$-channel projectors of irreducible
representations of $SU(3)$ 
\cite{Bartels}, and we have used explicitly $N_c=3$.

The soft anomalous dimension matrix in this basis is~\cite{KOS2}
\beq
\Gamma_{S'}=\left[\begin{array}{cc}
            \Gamma_{3 \times 3} & 0_{3 \times 5} \\
              0_{5 \times 3}      & \Gamma_{5 \times 5}
\end{array} \right] \, ,
\label{gammagggg}
\eeq
with
\beq
\blocA=\frac{\alpha_s}{\pi} \left[
                \begin{array}{ccc}
                  3\T  &   0  & 0  \\
                  0  &  3\U & 0    \\
                  0  &  0  &  3\left(\T+\U \right)
                   \end{array} \right]
\eeq
and
\beq
\Gamma_{5 \times 5}=\frac{\alpha_s}{\pi}\left[\begin{array}{ccccc}
6\T & 0 & -6\U & 0 & 0 \vspace{2mm} \\ 
0  & 3\T+\frac{3\U}{2} & -\frac{3\U}{2} & -3\U & 0 \vspace{2mm} \\ 
-\frac{3\U}{4} & -\frac{3\U}{2} &3\T+\frac{3\U}{2} & 0 & -\frac{9\U}{4} 
\vspace{2mm} \\
0 & -\frac{6\U}{5} & 0 & 3\U & -\frac{9\U}{5} \vspace{2mm} \\
0 & 0 &-\frac{2\U}{3} &-\frac{4\U}{3} & -2\T+4\U
\end{array} \right] \, .
\eeq
We note that the dependence on $\T$ is diagonal and in the forward region
of the partonic scattering, $\T \rightarrow -\infty$, 
where $\Gamma_{S}$ becomes diagonal, color singlet exchange dominates.
This has been a general trend for $\Gamma_S$ for all the processes
we analyzed in $t$-channel bases; suppression increases with the dimension
of the exchanged color representation.

The eigenvalues of the anomalous dimension matrix, 
Eq.~(\ref{gammagggg}), are
\beqa
\lambda_1&=&\lambda_4=3 \frac{\alpha_s}{\pi} \T, \quad
\lambda_2=\lambda_5=3 \frac{\alpha_s}{\pi} \U, \quad
\lambda_3=\lambda_6=3 \frac{\alpha_s}{\pi} (\T+\U), \nonumber\\
\lambda_{7,8}&=&2 \frac{\alpha_s}{\pi} \left[\T+\U \mp 2\sqrt{\T^2-\T\U+\U^2}
\right] \, .
\eeqa
The eigenvectors have the general form
\beq
e_i=\left[\begin{array}{c}
      e_i^{(3)} \\ 
       0^{(5)} 
\end{array}\right], \; i=1,2,3 \, , \; \; \; 
e_i=\left[\begin{array}{c}
     0^{(3)} \\ 
     e_i^{(5)} 
\end{array}\right], \; i=4 \ldots 8 \, ,
\eeq
where the superscripts refer to the dimension.
The three-dimensional vectors $e_i^{(3)}$ are defined by
\beq
e_i^{(3)}=\left[\begin{array}{c}
     \delta_{i1} \\ 
     \delta_{i2} \\  
        \delta_{i3} 
\end{array}\right], \; i=1,2,3 \, ,
\label{ei3}
\eeq
while $0^{(5)}$ and  $0^{(3)}$ are 
the five- and three-dimensional null vectors.
The five-dimensional vectors
$e_4^{(5)}$,  $e_5^{(5)}$ and 
$e_6^{(5)}$ are given by~\cite{KOS2}
\beq
e_4^{(5)}=\left[\begin{array}{c}
      -15 \vspace{2mm} \\ 
     6-\frac{15}{2}\frac{\T}{\U} \vspace{2mm} \\
      -\frac{15}{2}\frac{\T}{\U} \vspace{2mm} \\ 
        3 \vspace{2mm} \\ 
        1 \end{array} \right], \; \; \;
e_5^{(5)}=\left[\begin{array}{c}
      0   \vspace{2mm} \\
      -\frac{3}{2} \vspace{2mm} \\
       0  \vspace{2mm} \\
      \frac{3}{4}-\frac{3}{2}\frac{\T}{\U} \vspace{2mm}  \\
       1 \end{array} \right], \; \; \; 
e_6^{(5)}=\left[\begin{array}{c}
      -15  \vspace{2mm} \\
     -\frac{3}{2}+\frac{15}{2}\frac{\T}{\U} \vspace{2mm}  \\ 
      \frac{15}{2}-\frac{15}{2}\frac{\T}{\U} \vspace{2mm} \\ 
        -3 \vspace{2mm} \\ 
        1 \end{array} \right] \, .
\eeq
The expressions for $e_7^{(5)}$ and $e_8^{(5)}$ are long but can be given 
succinctly by~\cite{KOS2} 
\beqa
e_i^{(5)}&=&  \left[\begin{array}{c}
        b_1(\lambda_i') \vspace{2mm} \\
        b_2(\lambda_i') \vspace{2mm} \\ 
        b_3(\lambda_i') \vspace{2mm} \\
        b_4(\lambda_i') \vspace{2mm} \\
        1 
\end{array} \right], \; i=7,8 \, ,   
\eeqa
where 
\beq
\lambda_i'=\frac{\pi}{\alpha_s} \lambda_i \, ,
\eeq 
and where the $b_i$'s are given by
\beqa
b_1(\lambda_i')&=&\frac{3}{\U^2 K'} [80 \T^4+103 \U^4-280 \U \T^3 
-300 \T \U^3 +404 \T^2 \U^2
\nonumber \\ && \quad \quad
{}+(40 \T^3-16 \U^3 -60 \T^2 \U +52 \T\U^2)\lambda_i'] \, ,
\nonumber \\
b_2(\lambda_i')&=&\frac{3}{2K'}[20 \T^2-50 \U\T +44 \U^2
+(10 \T-5 \U)\lambda_i'] \, ,
\nonumber \\
b_3(\lambda_i')&=&-\frac{3}{2 \U K'}[40\T^3-64\U^3-120 \T^2 \U+130 \T \U^2
\nonumber \\ && \quad \quad
{}+(20 \T^2+13 \U^2-20 \T\U) \lambda_i'] \, ,
\nonumber \\
b_4(\lambda_i')&=&\frac{3\U}{K'}(2\T+5\U-2\lambda_i') \, ,
\eeqa
with
\beq
K'=20 \T^2-20 \U\T+21 \U^2 \, .
\eeq

With these results for the eigenvalues and eigenvectors we have all
the required elements for the diagonalization procedure.

\subsection{Diagonalization procedure for $gg \rightarrow gg$}

We now give a discussion of the diagonalization procedure
for the process $gg \rightarrow gg$. Since the anomalous dimension
matrix is $8 \times 8$ its diagonalization is quite more 
complicated than for the other processes that we have considered.

The Born cross section for $gg \rightarrow gg$
can be decomposed in the original basis,
Eq.~(\ref{8x8basis}), as
\beqa
\sigma^B&=&H_{11}|c_1|^2+H_{22}|c_2|^2+H_{33}|c_3|^2+H_{44}|c_4|^2 
\nonumber \\ &&
{}+H_{55}|c_5|^2+H_{66}|c_6|^2+H_{77}|c_7|^2+H_{88}|c_8|^2 \, .
\eeqa

Then if $C=(c_1 \; c_2 \; c_3 \; c_4 \; c_5 \; c_6 \; c_7 \; c_8)$
is the original color basis, the diagonal color basis is
\beq
C' \equiv (c_1' \; c_2' \; c_3' \; c_4' \; c_5' \; c_6' \; c_7' \; c_8')=CR\, ,
\eeq
where the columns of the matrix $R$ are the eigenvectors of $\Gamma_{S'}$:
\beq
R=\left[e_1 \; e_2 \; e_3 \; e_4 \; e_5 \; e_6 \; e_7 \; e_8 \right]
=\left[\begin{array}{cc}
1_{3 \times 3} & 0_{3 \times 5} \\
0_{5 \times 3} & R_{5 \times 5}
\end{array}\right] \, .
\eeq
Here 
\beq
R_{5 \times 5}=\left[\begin{array}{ccccc}
-15 & 0 & -15 & b_1(\lambda_7') & b_1(\lambda_8') \vspace{2mm} \\
6-\frac{15}{2}\frac{T}{U} & \frac{-3}{2} & \frac{-3}{2}+\frac{15}{2}\frac{T}{U}
& b_2(\lambda_7') & b_2(\lambda_8') \vspace{2mm} \\ 
-\frac{15}{2}\frac{T}{U} & 0 & \frac{15}{2}-\frac{15}{2}\frac{T}{U}
& b_3(\lambda_7') & b_3(\lambda_8') \vspace{2mm} \\
3 & \frac{3}{4}-\frac{3}{2}\frac{T}{U} & -3 & 
b_4(\lambda_7') & b_4(\lambda_8') \vspace{2mm} \\
1 & 1 & 1 & 1 & 1 
\end{array}\right] \, ,
\eeq
and the diagonalized anomalous dimension matrix is
$\Gamma_{S'}^{\rm diag}=R^{-1} \Gamma_{S'} R$.
To decompose the Born cross section in the diagonal basis we
use the relation $C=C' R^{-1}$.
The inverse of the matrix $R$ is
\beq
R^{-1}=\left[\begin{array}{cc}
1_{3 \times 3} & 0_{3 \times 5} \\
0_{5 \times 3} & R^{-1}_{5 \times 5}
\end{array}\right]
\eeq
where 
\beq
R^{-1}_{5 \times 5}=\left[\begin{array}{ccccc}
\frac{-U^2}{12 K_1} & \frac{4U^2-5TU}{15 K_1} & -\frac{UT}{3K_1} 
& \frac{U^2}{3 K_1} & \frac{3U^2}{20 K_1} \vspace{2mm} \\
0 & -\frac{16 U^2}{15 K_2} & 0 & \frac{4U^2-8TU}{3 K_2} 
& \frac{12 U^2}{5 K_2} \vspace{2mm} \\ 
-\frac{U^2}{12 K_3} & \frac{-U^2+5TU}{15 K_3} & \frac{U^2 -TU}{3K_3}
& -\frac{U^2}{3 K_3} & \frac{3U^2}{20 K_3} \vspace{2mm} \\
d_1(\lambda_7') & d_2(\lambda_7') & d_3(\lambda_7') & d_4(\lambda_7') 
& d_5(\lambda_7') \vspace{2mm} \\
-d_1(\lambda_8') & -d_2(\lambda_8') & -d_3(\lambda_8') & -d_4(\lambda_8') 
& -d_5(\lambda_8') \\
\end{array}\right] \, .
\eeq
Here we have simplified the expression for the inverse of $R$ by introducing
the variables $d$, which are given as functions of the eigenvalues 
$\lambda_i$, $i=7,8,$ by:
\beqa
d_1(\lambda_i')&=&U^2
\left[10T^3+15T^2U-19TU^2+24U^3 \right.
\nonumber \\ && \quad \quad 
\left. {}-\lambda_i'(10T^2-10TU+9U^2)\right]/(96 K_0 K_4) \, ,
\nonumber \\        
d_2(\lambda_i')&=&U^2 
\left[180T^5-360T^4U+509T^3U^2-339T^2U^3+166TU^4-6U^5 \right. 
\nonumber \\ && \hspace{-10mm} \left.
{}-\frac{\lambda_i'}{2}(60T^4-120T^3U+163T^2U^2-103TU^3+48U^4)\right]
/(12 K_0 K_5) \, ,
\nonumber \\   
d_3(\lambda_i')&=&U 
\left[30T^4-45T^3U+59T^2U^2-32TU^3+18U^4 \right.
\nonumber \\ &&   
\left. {}-\frac{\lambda_i'}{2}(10T^3-15T^2U+13TU^2-4U^3)\right]
/(12 K_0 K_4) \, , 
\nonumber \\
d_4(\lambda_i')&=&5U 
\left[240T^6-600T^5U+972T^4U^2-888T^3U^3+583T^2U^4 \right. 
\nonumber \\ && 
{}-205TU^5+48U^6 -\lambda_i\left(40T^5-100T^4U+152T^3U^2\right.
\nonumber \\ && \left. \left. 
{}-128T^2U^3+70TU^4-17U^5\right)\right]/(24 K_0 K_5) \, ,
\nonumber \\
d_5(\lambda_i')&=&
\left[2400T^7-7200T^6U+13440T^5U^2 
-15180T^4U^3+12306T^3U^4 \right.
\nonumber \\ && \hspace{-18mm}
{}-6321T^2U^5+2169TU^6-264U^7 
-\lambda_i \left(400T^6-1200T^5U+2140T^4U^2 \right.
\nonumber \\ && \hspace{-18mm} \left. \left.
{}-2280T^3U^3+1666T^2U^4-726TU^5+181U^6\right)\right]/(32 K_0 K_5) \, ,
\eeqa
where, again for brevity, we have introduced the notation 
\beqa
K_0&=&\sqrt{T^2-TU+U^2}, \quad K_1= 4U^2-4UT+5T^2,  
\nonumber \\ 
K_2&=&4T^2-4UT+5U^2, \quad K_3= 5U^2-6UT+5T^2,  
\nonumber \\ 
K_4&=&20U^4-44TU^3+69T^2U^2-50UT^3+25T^4,
\nonumber \\ 
K_5&=&100U^6-300U^5T+601U^4T^2-702U^3T^3+601T^4U^2
\nonumber \\ && \quad \quad 
-300T^5U+100T^6.
\eeqa

With these results we can now express the original basis in terms of the
new diagonal basis color tensors.
The  Born cross section can then be rewritten in the diagonal basis as
\beq
\sigma^B=\sum_{K,L=1}^8 \sigma^B_{KL} \, ,
\eeq
where $\sigma^B_{KL}$ is the component of the Born cross section
proportional to $c_K' c_L'^*$.
While the calculation of these components is straightforward,
the explicit expressions are quite long.

The resummed cross section is then given by
\beq
\sigma^{\rm res}=\sum_{K,L=1}^8 \sigma^B_{KL} \, e^{E_{KL}} \, ,
\eeq
where $E_{KL}$ takes contributions at NLL from the eigenvalues
$\lambda_K$ and $\lambda_L$.

\mysection{Threshold resummation for direct photon and
$W$ boson production}

\subsection{Factorization and resummed cross sections}

In this section we discuss resummation for direct photon and
electroweak boson production. Threshold resummation may be important
for these processes at large transverse momentum.
Here we formulate the resummation in single-particle inclusive 
kinematics~\cite{LOS}.
Resummation will follow, as we saw before, from the factorization
properties of the cross section.

\begin{figure}
\centerline{
\psfig{file=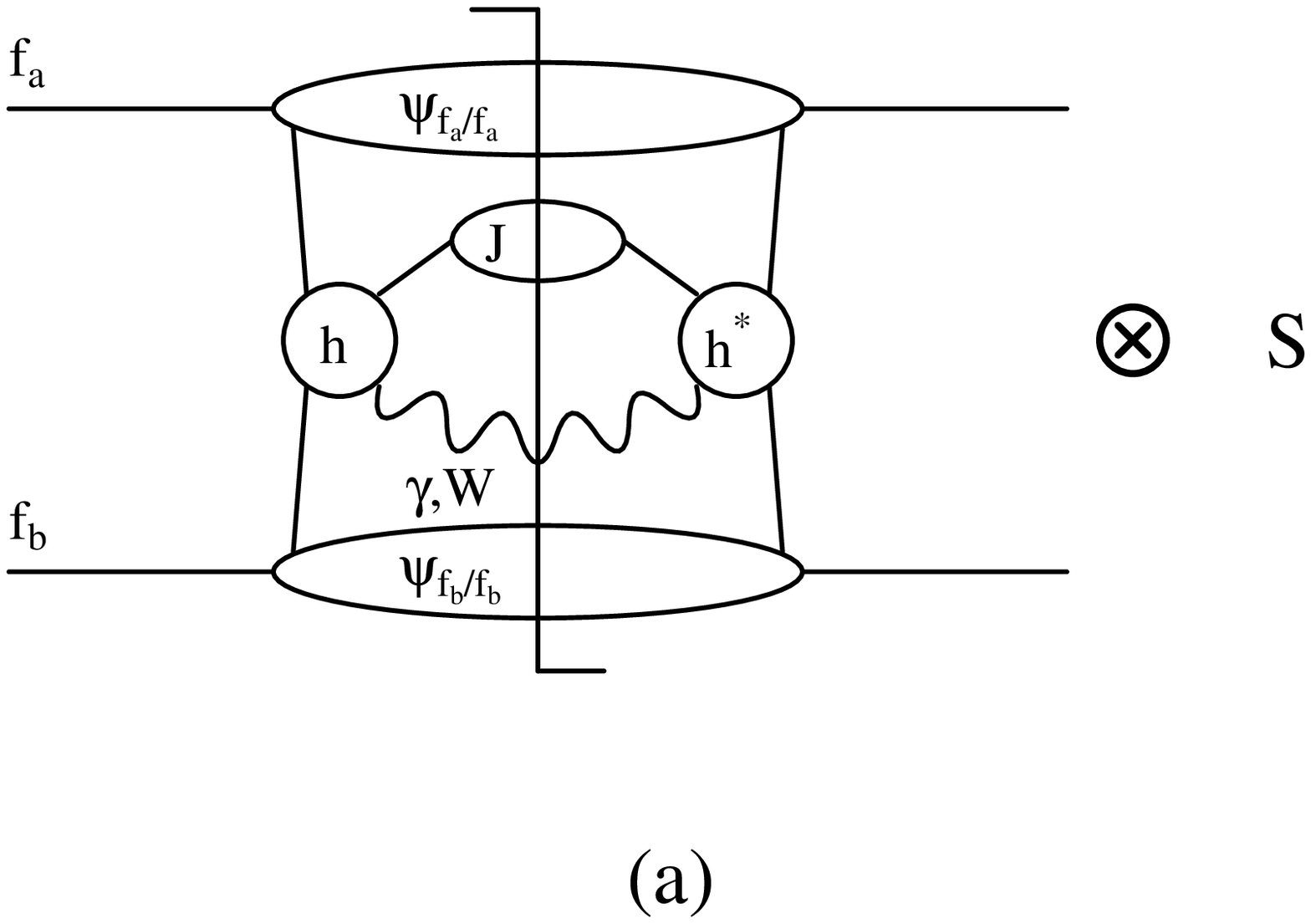,height=2.5in,width=4.05in,clip=}}
\vspace{5mm}
\centerline{
\psfig{file=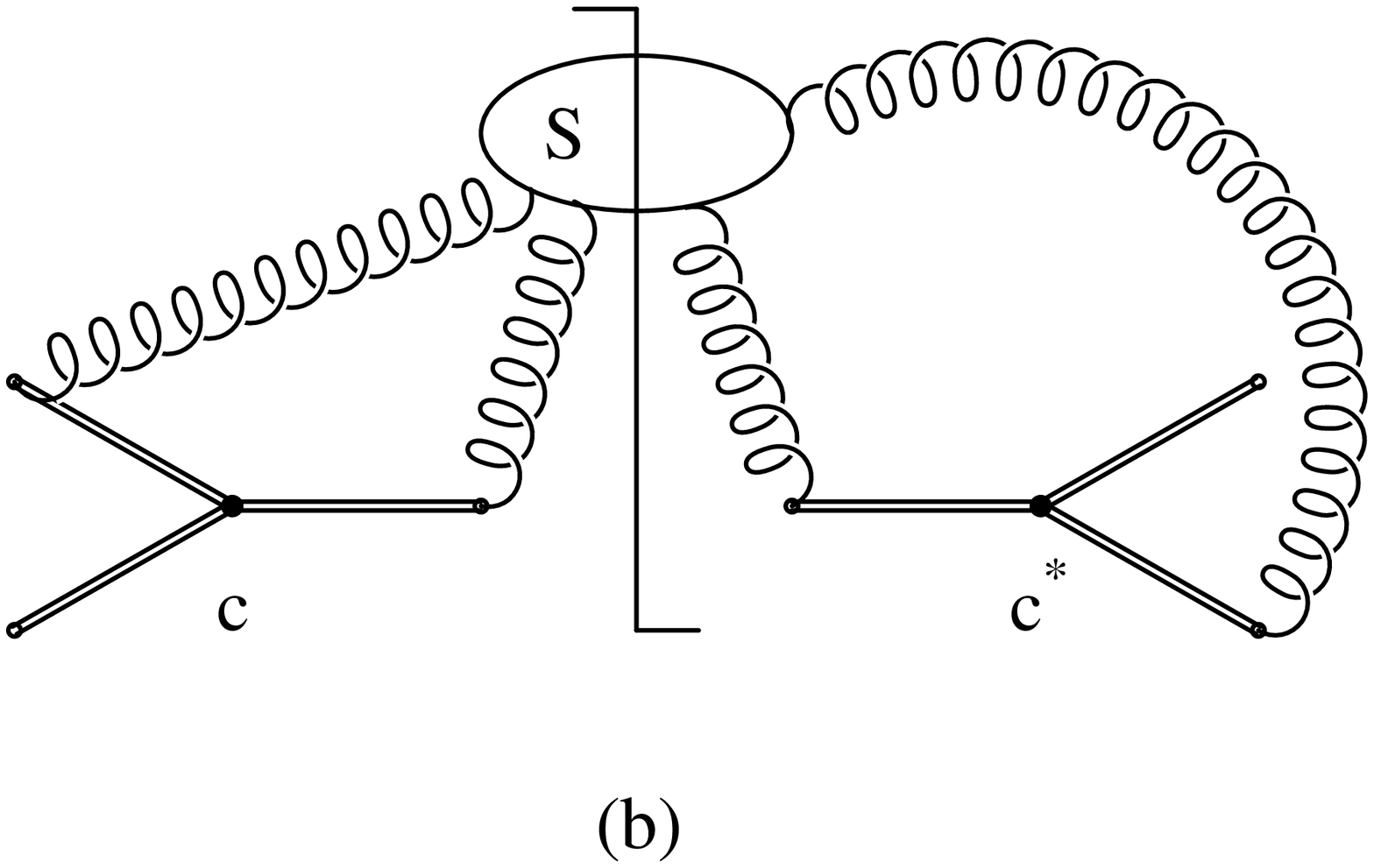,height=2.5in,width=4.05in,clip=}}
{Fig. 7. (a) Factorization for direct photon or $W$ + jet production 
near partonic threshold.
(b) The soft-gluon function $S$, in which
the vertices $c$ link ordered exponentials.}
\label{fig7}
\end{figure}

We consider hadronic processes of the form
\beq
h_A(p_A)+h_B(p_B) \rightarrow \gamma,W(Q) +X \, .
\eeq
The partonic subprocesses contributing to direct photon or $W$ + jet
production are
\beq
q(p_a)+g(p_b) \longrightarrow q(p_J) + \gamma \, , W(Q)
\eeq
and
\beq
q(p_a)+{\bar q}(p_b) \longrightarrow g(p_J) + \gamma \, , W(Q) \, .
\eeq
We define Mandelstam invariants
\beq
s=(p_a+p_b)^2 \, , \quad t=(p_a-Q)^2 \, , 
\quad u=(p_b-Q)^2 \, ,
\eeq
which satisfy $s+t+u=Q^2$ at threshold, with $Q^2=M_W^2$ for $W$ production 
and $Q^2=0$ for direct photon production.

The factorized form of the cross section for direct photon 
or $W$ + jet production is a convolution of the parton distribution functions
$\phi$ with the hard scattering function, $\hat{\sigma}$:
\beqa
E_Q\frac{d\sigma_{h_Ah_B\rightarrow\gamma,W}}{d^3 Q}&=&
\sum_{ab}\int dx_a dx_b \,  \phi_{f_a/h_A}(x_a,\mu^2) \,
\phi_{f_b/h_B}(x_b,\mu^2) 
\nonumber \\ && \hspace{20mm}
\times \; {\hat \sigma}(s_2,t,u,Q^2,\alpha_s(\mu^2)) \, ,
\eeqa
where we define $s_2=s+t+u-Q^2$, with $Q=p_W$ for $W$ production
and $Q=p_{\gamma}$ for direct photon production.
The threshold region is given by $s_2=0$.

The cross section can be refactorized into distributions $\psi$,
defined in direct analogy with the c.m. distributions that we presented
in Section 2, hard components $H=h^*h$, and a soft gluon 
function $S$. This refactorization, similar to the ones for heavy quark 
and dijet production, is shown in Fig. 7. There are some differences
in the details of the formalism here as compared to the results in the 
previous sections because the resummation is now carried out 
in single-particle inclusive kinematics.
More details are given in Ref.~\cite{LOS}.   

The color structure of the hard scattering for direct photon or
W boson production is much simpler 
than for heavy quark or dijet production. The color basis here
consists of only one tensor; hence $\Gamma_S$ is simply a
$1 \times 1$ matrix, and no color traces or path ordering appear 
in the resummed expressions.

The resummation of the $N$-dependence of each of the functions in the
refactorized cross section leads to the expression  in 
the $\overline{\rm MS}$ factorization scheme in single-particle 
inclusive kinematics:
\beqa
\tilde{{\sigma}}(N) &=&  
\exp \left \{ \sum_{i=a,b} \left [E^{(f_i)}(N_i,p_i \cdot \zeta)\right.\right. 
\nonumber\\ && \hspace{20mm} \left. \left. 
{}-2\int_\mu^{2 p_i \cdot \zeta}{d\mu'\over\mu'}\; 
\left [\gamma_{f_i}(\alpha_s(\mu'{}^2))-\gamma_{f_if_i}
(N_i,\alpha_s(\mu'{}^2)) 
\right] \right] \right\}
\nonumber \\ && \times \; 
\exp \left \{E'_{(f_J)}(N,p_J \cdot n) \right\} \;
H\left(t,u,Q^2,\alpha_s(\mu^2)\right) 
\nonumber \\ && \hspace{-10mm} \times \; \left.  
S\left(1,\beta_i,\zeta, n, \alpha_s(S/N^2)\right) \;
\exp \left[\int_\mu^{\sqrt{S}/N} {d\mu' \over \mu'} \, 
2 \, {\rm Re} \Gamma_S\left(\alpha_s(\mu'^2)\right)\right]
\right\} \, ,
\nonumber \\
\eeqa
where $\zeta^{\mu}=p_3^{\mu}/\sqrt{S}$~\cite{LOS} and $\beta_i$ are
the particle velocities. 
The first exponent $E^{(f_i)}$ has the same form as in Eq.~(\ref{Eexp})
with $N_a=N(-u/s)$ and  $N_b=N(-t/s)$.
The exponent $E'_{(f_J)}$ has the same form as in Eq.~(\ref{Eprexp}),
with $A^{(f)}$ given by (\ref{Aexp}) and $B'_{(f)}$ given for quarks 
by~\cite{LOS}
\beq
B'_{(q)}=\frac{\alpha_s}{\pi}C_F\left[-\frac{7}{4}+\ln(2\nu_q)\right]
\eeq
and for gluons by
\beq
B'_{(g)}=\frac{\alpha_s}{\pi}C_A\left[\frac{n_f}{6C_A}-\frac{11}{12}-1
+\ln(2\nu_g)\right] \, .
\eeq

\subsection{Soft anomalous dimensions and expansions of 
\protect\newline the resummed cross sections}

Here we give the anomalous dimensions for direct photon 
and $W$ + jet production. 
The calculation follows the same lines as for heavy quark and
dijet production. The main difference here is that the color basis
consists of  only one tensor, $c=T_F$, and there are three eikonal
lines connecting at the color vertex. The one-loop eikonal vertex 
corrections for the partonic subprocesses in direct photon and 
$W$ + jet production are shown in Fig. 8. 

\begin{figure}
\centerline{
\psfig{file=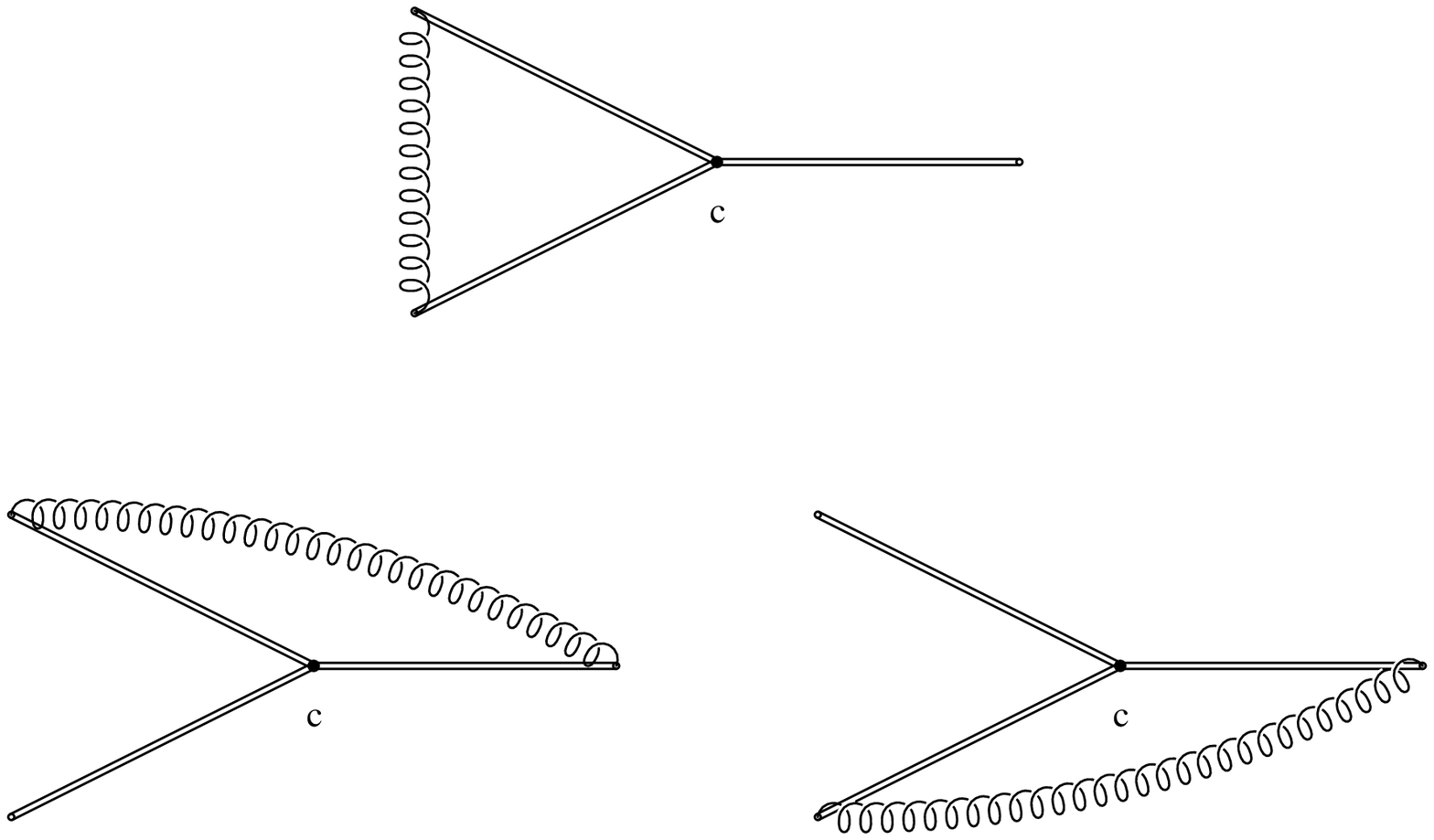,height=2.5in,width=4.05in,clip=}}
{Fig. 8. One-loop eikonal vertex corrections for 
partonic subprocesses in direct photon and $W$ + jet production.}
\label{fig8}
\end{figure}

The anomalous dimension for the process $q(p_a)+ g(p_b)  
\rightarrow q(p_J)+\gamma(Q)$ 
or the process $q(p_a)+g(p_b) \rightarrow q(p_J)
+W(Q)$ is given by
\beqa
\Gamma_S&=&\frac{\alpha_s}{2\pi}\left\{C_F\left[2\ln\left(\frac{-u}{s}\right)
-\ln(4\nu_{q_a} \nu_{q_J})+2 \right] \right.
\nonumber \\ &&  \quad \quad \left.
{}+C_A\left[\ln\left(\frac{t}{u}\right)
-\ln(2 \nu_g) +1 -\pi i \right]\right\} \, .
\eeqa

The one-loop expansion of the resummed cross section 
for direct-photon production is given in 
Ref.~\cite{LOS}. The authors have found agreement with the exact NLO
results in Ref.~\cite{dgamma1loop}.  
Numerical results are given in Ref.~\cite{NKJO}.
We may also expand the resummed cross section for $W$ + jet production.
The $\overline {\rm MS}$ one-loop expansion at NLL of $s_2/Q^2$
in single-particle inclusive kinematics is
\beqa   
\hspace{-15mm}
{\hat \sigma}^{\overline {\rm MS}\, (1)}_{qg \rightarrow qW}(s_2,s,t,u,Q^2)
&=&\sigma^B_{qg\rightarrow qW} \frac{\alpha_s}{\pi} \left\{ 
(C_F+2C_A) \left[\frac{\ln(s_2/Q^2)}{s_2}\right]_+ \right.
\nonumber \\ &&  \hspace{-45mm} \left.
{}-(C_F+C_A)\ln\left(\frac{\mu^2}{Q^2}\right)
\left[\frac{1}{s_2}\right]_+ 
+\left[-\frac{3}{4}C_F+C_A \ln \left(\frac{sQ^2}{tu}\right)\right]
\left[\frac{1}{s_2}\right]_+ \right\} ,
\eeqa
where $\sigma^B_{qg\rightarrow qW}$ denotes the Born cross section for this
partonic channel.
This result agrees with the exact NLO expressions in Ref. \cite{wjet}.

The anomalous dimension for the process $q(p_a)+{\bar q}(p_b) 
\rightarrow g(p_J)+ \gamma(Q)$
or the process $q(p_a)+{\bar q}(p_b) \rightarrow g(p_J)+W(Q)$ 
is given by
\beqa
\Gamma_S&=&\frac{\alpha_s}{2\pi}\left\{C_F\left[
-\ln(4\nu_q \nu_{\bar q})+2 -2 \pi i\right] \right.
\nonumber \\ && \quad \quad \left. 
{}+C_A\left[\ln\left(\frac{tu}{s^2}\right)
-\ln(2 \nu_g) +1 +\pi i \right]\right\} \, .
\eeqa

Again, the one-loop expansion for direct-photon production is given in 
Ref.~\cite{LOS} and agreement has been found with the exact NLO
results in \cite{dgamma1loop}.  
The $\overline {\rm MS}$ one-loop expansion at NLL of $s_2/Q^2$
for $W$ + jet production in this channel and in single-particle 
inclusive kinematics is 
\beqa
{\hat \sigma}^{\overline {\rm MS} \, (1)}_{q{\bar q} 
\rightarrow gW}(s_2,s,t,u,Q^2)
&=&\sigma^B_{q{\bar q}\rightarrow gW} \frac{\alpha_s}{\pi} \left\{ 
(4C_F-C_A) \left[\frac{\ln(s_2/Q^2)}{s_2}\right]_+ \right. 
\nonumber \\ && \hspace{-25mm} 
{}-2 C_F \ln\left(\frac{\mu^2}{Q^2}\right)
\left[\frac{1}{s_2}\right]_+ 
+\left[2C_F\ln \left(\frac{sQ^2}{tu}\right) \right.
\nonumber \\ && \hspace{-20mm} \left. \left.
{}+C_A \left(\frac{n_f}{6C_A}-\frac{11}{12}
-\ln \left(\frac{sQ^2}{tu}\right)\right)\right]
\left[\frac{1}{s_2}\right]_+ \right\} \, ,
\eeqa
which again agrees with the exact NLO results in Ref. \cite{wjet}.

One may of course expand the resummed cross sections to two loops or higher 
for both direct photon~\cite{NKJO} and $W$ boson~\cite{NKVD} production.
For example, the $\overline {\rm MS}$ two-loop expansion at NLL 
of $1-w=s_2/(s+t)$ for direct photon production in the channel 
$q {\bar q} \rightarrow g \gamma$ is
\beqa
{\hat \sigma}^{\overline {\rm MS} \, (2)}_{q{\bar q} 
\rightarrow g \gamma}(1-w,s,v)
&=&\sigma^B_{q{\bar q}\rightarrow g \gamma} \frac{\alpha_s^2}{\pi^2} 
\left\{\left(8C_F^2-4C_FC_A+\frac{C_A^2}{2}\right) 
\left[\frac{\ln^3(1-w)}{1-w}\right]_+ \right. 
\nonumber \\ && \hspace{-40mm} 
{}+\left[-12C_F^2\left(\ln\left(\frac{1-v}{v}\right)
+\ln\left(\frac{\mu^2}{s}\right)\right)
+C_A^2\left(-\frac{3}{2}\ln(1-v)-\frac{n_f}{4C_A}+\frac{11}{8}\right)\right.
\nonumber \\ && \hspace{-30mm}
{}+C_FC_A\left(9\ln(1-v)-3\ln v+\frac{n_f}{C_A}-\frac{11}{2}
+3\ln\left(\frac{\mu^2}{s}\right)\right)
\nonumber \\ && \hspace{-30mm} \left. \left.
{}-\beta_0\left(C_F-\frac{3}{8}C_A\right)\right]
\left[\frac{\ln^2(1-w)}{1-w}\right]_+ \right\} \, ,
\eeqa
where $v=1+t/s$.

\mysection{Conclusion}

We have reviewed the resummation of threshold logarithms for
heavy quark, dijet, direct photon, and $W$ boson production
in hadronic collisions. Resummation follows from the factorization
properties of the cross sections and is influenced by the
color exchange in the hard scattering.
We have presented full results for the soft anomalous dimension matrices
for all the relevant partonic subprocesses. The one-loop 
expansions of the resummed cross sections agree with exact
NLO calculations and one can expand the cross sections to two-loops
or higher orders. We have discussed the general diagonalization 
procedure that can be implemented in the calculation of the resummed
cross sections. Numerical results have been presented for 
top quark production at the Fermilab Tevatron. 
The resummation formalism is quite general and can be applied as well to
the calculation of transverse momentum or other differential distributions 
for a variety of processes, such as heavy quark production. 

\mysection*{Acknowledgements}

This work was supported in part by the U.S. Department of Energy.
I wish to thank Gianluca Oderda, Jack Smith, George Sterman, and 
Ramona Vogt for very productive past and present collaborations. 
I would also like to thank Vittorio Del Duca, Eric Laenen, Sven Moch, 
and Jeff Owens for many useful conversations and current collaborations. 

\appendix
\section{Evaluation of one-loop eikonal vertex 
\protect\newline corrections}
\label{app-integ}

Here we give some calculational details for the one-loop 
corrections of Figs. 2 and 8.

In our calculations we use Feynman eikonal rules as discussed in
Sections 3 and 4, and a general axial gauge gluon propagator,
\begin{equation}
D^{\mu \nu}(k)=\frac{-i}{k^2+i\epsilon} N^{\mu \nu}(k), \quad
N^{\mu \nu}(k)=g^{\mu \nu}-\frac{n^{\mu}k^{\nu}+k^{\mu}n^{\nu}}{n \cdot k}
+n^2\frac{k^{\mu}k^{\nu}}{(n \cdot k)^2},
\end{equation}
with $n^{\mu}$ the axial gauge-fixing vector.

We denote the kinematic part of the one-loop vertex correction to 
$c_I$, with the virtual gluon linking lines $v_{i}$ and $v_{j}$,
as $\omega_{ij}(\delta_{i}v_{i},\delta_{j}v_{j},\Delta_{i},\Delta_{j})$.
The $\delta$'s and $\Delta$'s are defined as in Sections 3 and 4 except 
that here we use $\Delta=-i$ $(+i)$ for a gluon located below (above) the
eikonal line in order to present the results 
for both quarks and gluons in a uniform fashion.

The expression for $\omega_{ij}$ is then
\beqa   
\omega_{ij}(\delta_{i}v_{i},\delta_{j}v_{j},\Delta_{i},\Delta_{j})&=&
{g}_{s}^2\int\frac{d^nq}{(2\pi)^n}\frac{-i}{q^2+i\epsilon}
\left\{\frac{\Delta_{i} \: \Delta_{j} \:v_{i}{\cdot}v_{j}}{(\delta_{i}v_{i}
{\cdot}q+i\epsilon)
(\delta_{j}v_{j}{\cdot}q+i\epsilon)}\right. 
\nonumber\\ &&
\hspace{-35mm} \left.{}-\frac{\Delta_{i}v_{i}{\cdot}n}{(\delta_{i}v_{i}
{\cdot}q+i\epsilon)}
\frac{P}{(n{\cdot}q)}-\frac{\Delta_{j}v_{j}{\cdot}n}{(\delta_{j}v_{j}{\cdot}q
+i\epsilon)}
\frac{P}{(n{\cdot}q)}+n^2\frac{P}{(n{\cdot}q)^2}\right\},         
\label{omega}
\eeqa
with $g_s^2=4\pi \alpha_s$, and where $P$ stands for principal value,
\beqa
\frac{P}{(q \cdot n)^{\beta}}=\frac{1}{2}\left(\frac{1}{(q \cdot n+i\epsilon)
^{\beta}}+(-1)^{\beta}\frac{1}{(-q \cdot n+i\epsilon)^{\beta}}\right).
\eeqa
We may rewrite (\ref{omega}) as
\beqa  
\omega_{ij}(\delta_{i}v_{i},\delta_{j}v_{j},\Delta_{i},\Delta_{j})&=& 
{\cal S}_{ij}
\left[I_1(\delta_i v_i, \delta_j v_j)
-\frac{1}{2}I_2(\delta_i v_i, n)-\frac{1}{2}I_2(\delta_i v_i, -n)\right.
\nonumber \\ && 
\left.{}-\frac{1}{2}I_3(\delta_j v_j, n)-\frac{1}{2}I_3(\delta_j v_j, -n)
+I_4(n^2)\right] \, ,
\label{omegaI}
\end{eqnarray}
where ${\cal S}_{ij}$ is an  overall sign
\beq
{\cal S}_{ij}=\Delta_i \: \Delta_j \: \delta_i \: \delta_j.
\eeq

We now evaluate the ultraviolet poles of the integrals.
For the integrals when both $v_i$ and $v_j$ refer to massive quarks we have
(with $\epsilon=4-n$) \cite{Thesis,KS}
\begin{eqnarray}
I_1^{\rm{UV \;pole}}&=&\frac{\alpha_s}{\pi}\frac{1}{\epsilon} L_{\beta} \, ,
\nonumber\\
I_2^{\rm{UV \;pole}}&=&-\frac{\alpha_s}{\pi}\frac{1}{\epsilon} L_i \, ,
\nonumber\\
I_3^{\rm{UV \;pole}}&=&-\frac{\alpha_s}{\pi}\frac{1}{\epsilon} L_j \, ,
\nonumber\\
I_4^{\rm{UV \;pole}}&=&-\frac{\alpha_s}{\pi}\frac{1}{\epsilon} \, ,
\end{eqnarray}
where the $L_\beta$ is the  velocity-dependent
eikonal function
\begin{equation}
L_{\beta}=\frac{1-2m^2/s}{\beta}\left(\ln\frac{1-\beta}{1+\beta}
+\pi i \right)\, ,
\end{equation}
with $\beta=\sqrt{1-4m^2/s}$.
The $L_i$ and $L_j$ are rather complicated functions of
the gauge vector $n$. Their
contributions are cancelled by the inclusion of self energies.
Their explicit expressions are:
\begin{equation}
L_i=\frac{1}{2}\, [L_i(+n)+L_i(-n)] \, ,
\label{Ellidef}
\end{equation}
where
\begin{eqnarray}
L_i(\pm n)&=&\frac{1}{2}\frac{|v_i \cdot n|}{\sqrt{(v_i \cdot n)^2-2m^2n^2/s}}
\nonumber \\ && \hspace{-5mm}  \times \, 
\left[\ln\left(\frac{\delta(\pm n) \, 2m^2/s-|v_i \cdot n|
- \sqrt{(v_i \cdot n)^2-2m^2n^2/s}}
{\delta(\pm n) \, 2m^2/s-|v_i \cdot n|
+ \sqrt{(v_i \cdot n)^2-2m^2n^2/s}}\right)\right.
\nonumber \\ &&
{}+\left.\ln\left(\frac{\delta(\pm n) \, n^2-|v_i \cdot n|
- \sqrt{(v_i \cdot n)^2-2m^2n^2/s}}
{\delta(\pm n) \, n^2-|v_i \cdot n|
+ \sqrt{(v_i \cdot n)^2-2m^2n^2/s}}\right)\right]
\end{eqnarray}
with $\delta(n) \equiv |v_i \cdot n|/ (v_i \cdot n)$.
Then, using Eq. (\ref{omegaI}), we get
\beq       
\omega_{ij}(\delta_{i}v_{i},\delta_{j}v_{j},\Delta_{i},\Delta_{j})=
{\cal S}_{ij} \, \frac{\alpha_{s}}{\pi\epsilon}
\left[L_{\beta} + L_i + L_j -1 \right].
\label{omegaheavy}
\eeq
The contribution of the heavy quark self energy graphs in Fig.~2(b)
to the anomalous dimension matrices for heavy quark production
in either partonic channel is $(\alpha_s/\pi)C_F(L_1+L_2-2) \delta_{IJ}$,
which cancels the gauge-dependent contribution from the $\omega_{12}$
in Eq.~(\ref{omegaheavy}).  

When $v_i$ refers to a massive quark and $v_j$ to a massless quark 
we have \cite{Thesis,KS}
\begin{eqnarray}
I_1^{\rm{UV \;pole}}&=&\frac{\alpha_s}{2\pi}
\left\{\frac{2}{\epsilon^2}-\frac{1}{\epsilon}
\left[\gamma+\ln\left(\frac{v_{ij}^2 s}{2m^2}\right)-\ln(4\pi) \right]
\right\}\, , \nonumber\\
I_2^{\rm{UV \;pole}}&=&-\frac{\alpha_s}{\pi}\frac{1}{\epsilon} L_i \, ,
\nonumber\\
I_3^{\rm{UV \;pole}}&=&\frac{\alpha_s}{2\pi}
\left\{\frac{2}{\epsilon^2}-\frac{1}{\epsilon}
\left[\gamma+\ln \nu_j-\ln(4\pi)\right]
\right\} \, ,
\nonumber\\
I_4^{\rm{UV \;pole}}&=&-\frac{\alpha_s}{\pi}\frac{1}{\epsilon} \, ,
\end{eqnarray}
where
\begin{equation}
\nu_a=\frac{(v_a \cdot n)^2}{|n|^2}\, ,
\end{equation}
and $v_{ij}=v_i \cdot v_j$. Note that the double poles cancel
in the sum over the $I$'s and we get
\beq       
\omega_{ij}(\delta_{i}v_{i},\delta_{j}v_{j},\Delta_{i},\Delta_{j})=
{\cal S}_{ij} \, \frac{\alpha_{s}}{\pi\epsilon}
\left[-\frac{1}{2}\ln\left(\frac{v_{ij}^2s}{2m^2}\right) + L_i 
+\frac{1}{2}\ln \nu_{j} -1\right].
\eeq

Finally, when both $v_i$ and $v_j$ refer to massless quarks we have
\cite{BottsSt,Thesis,KS}
\begin{eqnarray}
I_1^{\rm{UV \;pole}}&=&\frac{\alpha_s}{\pi}
\left\{\frac{2}{\epsilon^2}-\frac{1}{\epsilon}
\left[\gamma+\ln\left(\delta_i\delta_j\; \frac{v_{ij}}{2}\right)
-\ln(4\pi) \right]\right\} \, ,
\nonumber\\
I_2^{\rm{UV \;pole}}&=&\frac{\alpha_s}{2\pi}
\left\{\frac{2}{\epsilon^2}-\frac{1}{\epsilon}
\left[\gamma+\ln \nu_i -\ln(4\pi)\right]\right\} \, ,
\nonumber\\
I_3^{\rm{UV \;pole}}&=&\frac{\alpha_s}{2\pi}
\left\{\frac{2}{\epsilon^2}-\frac{1}{\epsilon}
\left[\gamma+\ln \nu_j -\ln(4\pi)\right]\right\} \, ,
\nonumber\\
I_4^{\rm{UV \;pole}}&=&-\frac{\alpha_s}{\pi}\frac{1}{\epsilon}\, .
\end{eqnarray}
Again, note that the double poles cancel in the sum over the $I$'s and we get
\beq
\omega_{ij}(\delta_{i}v_{i},\delta_{j}v_{j},\Delta_{i},\Delta_{j})=
{\cal S}_{ij} \, \frac{\alpha_{s}}{\pi\epsilon}
\left[-\ln\left(\delta_{i} \, \delta_{j} \, 
\frac{v_{ij}}{2}\right)+\frac{1}{2}\ln(\nu_{i}\nu_{j})-1\right].
\eeq
In order to obtain contributions to the different entries of the matrix of 
renormalization constants,
the above expression has still to be multiplied by the color decomposition of 
its corresponding 
diagram into the basis color structures \cite{Thesis,KS,BottsSt}.

\end{document}